\documentclass[journal,10pt,doublecolumn]{IEEEtran}
\ifCLASSINFOpdf
   \usepackage[pdftex]{graphicx}
   \usepackage{rotating}
   \usepackage{subcaption}
   \usepackage[dvipsnames,table]{xcolor}
     \usepackage{amsmath}
   \usepackage{amsfonts}
  \usepackage{wasysym}
  \usepackage{multirow}
  \usepackage{balance}

\else
\fi
\begin{document}
%
\title{Predictive No-Reference Assessment of Video Quality}
%
%
%

\author{Maria~Torres~Vega,~\IEEEmembership{Student Member,~IEEE,}
	Decebal~Constantin~Mocanu,~\IEEEmembership{Student Member,~IEEE,}
        and~Antonio~Liotta,~\IEEEmembership{Senior Member,~IEEE}
\thanks{M. Torres Vega, D.C Mocanu and A. Liotta are with the Department of Electrical Engineering, Eindhoven University of Technology, P.O. Box 513, 5600MB, Eindhoven, the Netherlands, e-mails: \{m.torres.vega, d.c.mocanu, a.liotta\}@tue.nl}
}

%
%

\markboth{THIS ARTICLE IS A PRE-PRINT VERSION AND CURRENTLY UNDER REVIEW.}%
{Torres Vega\MakeLowercase{\textit{et al.}}: Predicted No-Reference Assessment of Video Quality}
%



\maketitle

\begin{abstract}
Among the various means to evaluate the quality of video streams, No-Reference (NR) methods have low computation and may be executed on thin clients. Thus, NR algorithms would be perfect candidates in cases of real-time quality assessment, automated quality control and, particularly, in adaptive mobile streaming. Yet, existing NR approaches are often inaccurate, in comparison to Full-Reference (FR) algorithms, especially under lossy network conditions. In this work, we present an NR method that combines machine learning with simple NR metrics to achieve a quality index comparably as accurate as the Video Quality Metric (VQM) Full-Reference algorithm. Our method is tested in an extensive dataset (960 videos), under lossy network conditions and considering nine different machine learning algorithms. Overall, we achieve an over 97\% correlation with VQM, while allowing real-time assessment of video quality of experience in realistic streaming scenarios. 
\end{abstract}

\begin{IEEEkeywords}
Quality of Experience, No-Reference Video Quality Assessment, Supervised Machine Learning.
\end{IEEEkeywords}

%
\IEEEpeerreviewmaketitle

\section{Introduction}
\label{sec:introduction}

No-Reference (NR) video quality methods have the potential to provide real-time video quality assessment and automated quality control, for instance in the context of mobile streaming. This is because NR algorithms are computationally light and do not require comparing the video stream under scrutiny with its original (unimpaired) benchmark, as would be the case of Full-Reference(FR) methods~\cite{journals/tbc/ChikkerurSRK11}.

Due to their particular methodology, computational requirements and functional limitations, neither FR methods nor subjective evaluations are viable to automate quality control processes. Subjective studies are performed off-line but are instrumental in understanding quality perception, i.e. Quality of Experience (QoE)~\cite{qualinet}~\cite{Varela2015QoE}. On the other hand, FR algorithms such as the Video Quality Metric (VQM)~\cite{vqm} have proven to correlate well with the human vision system~\cite{journals/tbc/ChikkerurSRK11} and this is the reason why many studies use them to benchmark other simpler algorithms, rather than being used directly in video management applications~\cite{Pandremmenou2015}.

This is in fact the approach we use in our work, where we aim to introduce a new NR method that combines the simplicity (and applicability) of NR metrics with the accuracy that is typically achieved only through heavyweight FR methods. In this way, we make a breakthrough in video quality analysis, enabling a whole new range of applications which would not be functional with the current NR methods, due to their proven inaccuracy, particularly under lossy network conditions~\cite{TorresES2016}.

Nowadays, Video Quality Assessment (VQA) methods and metrics are drawn from knowledge in human QoE and perception~\cite{NRReviewShahid2014}. At its essence, VQA is a subjective matter, best judged by human subjects, as in subjective studies and subjective analyses~\cite{journals/tbc/Huynh-ThuGSCR11}. Typically, sample people (chosen from different representative categories) rate video quality (or quality variations), under controlled conditions, following well-established methods~\cite{conf2010-Hossfeld}. The outcomes are given in terms of Mean Opinion Score (MOS) or any other derived metric. Although well-aligned to human perception, subjective studies are costly, time-consuming and prone to human bias. They are fundamental to the various applications of VQA, yet great effort has been directed towards mimicking subjective studies through completely automated processes and algorithms, as in objective QoE~\cite{info3-conf-2008-303}.

Traditionally, objective methods use as input the original reference signal (e.g. image, video, audio) and a distorted version. In our context, this will be a video sequence distorted by compression and network impairments. FR QoE aims to estimate the perceptual degradation in the distorted sequence, compared to the reference sequence~\cite{journals/tbc/ChikkerurSRK11}. Perhaps the simplest, most popular and less accurate among FR algorithms is the Peak Signal to Noise Ratio (PSNR)~\cite{2008winkler_video_quality_metrics_evolution}, derived directly from the Mean Square Error~\cite{lehmann1998}. A better compromise between complexity and accuracy is offered by the Structural Similarity (SSIM)~\cite{ssimmetric1,ssimmetric2}, which combines video luminance, contrast and structure to evaluate the quality degradation at frame-by-frame level. When the inter-frame degradations are of interest (for instance in the presence of network-impaired video streams), VQM is a better option~\cite{journals/tbc/ChikkerurSRK11}.

Although not perfectly, FR metrics provide the best correlation with human perception, but are not always applicable in real systems due to the requirements to have both the reference and the distorted sequence available. Also the more accurate FR metrics are computationally demanding and are, instead, more effective to generate offline benchmarking, as we do in our study.

To the other end of the spectrum, stand the NR~\cite{NRReviewShahid2014} metrics, that operate merely on the distorted sequence (e.g. the video stream rendered after network transmission, as in our case). These metrics are algorithmically simple, since they focus on specific features~\cite{journals/tip/CiancioCSSSO11, BorerJ2014, 5140}, which are only indicative of quality and do not always correlate well with subjective or FR results. In previous research, we analyzed a range of state-of-the-art NR metrics on a large video dataset and packet losses in a 0-10\% range~\cite{TorresES2016}. We showed how different metrics capture diverse types of distortions, concluding that none of the existing metrics is universally effective. Also, all metrics failed under lossy networks. 

Given the complexity of FR methods and the inaccuracy of NR methods, the aim of this paper is to explore how Machine Learning (ML) may lead to an accurate NR method. This is a new direction in the development of NR algorithms. Promising examples are the bitstream based artificial neural network of Shahid et al.~\cite{DBLP:conf/qomex/ShahidPWBL15}, the artificial neural network for jerkiness evaluation~\cite{DBLP:journals/corr/XueEW14}, the Least Absolute Shrinkage and Selection Operator (LASSO) algorithm~\cite{Pandremmenou2015}, which were evaluated on synthetically impaired videos. Yet, our aim is to find a method that can work in general cases, particularly in real-time streaming and including realistic network impairments. Our method analyses the received video stream in terms of eight NR features (both on the bitstream and the pixel levels) in addition to sensing the network to obtain two network measurements (nominal bitrate and estimated level of packet loss). These ten features serve as input to a Supervised Learning (SL) algorithm that, based on previously learned samples of video quality, performs a predictive assessment of the quality of the video under scrutiny.

We extensively tested our method in a large self-developed video impaired dataset (derived from ten original videos of the Live Video Database~\cite{Seshadrinathan:2010:SSO:1827592.1827596}), considering nine different types of SL algorithms and using VQM as the benchmark quality assessment method. We achieved high accuracy, obtaining an overall correlation to VQM higher than $97\%$. In this paper, we present the method and its evaluation. As an additional contribution, we provide a comparative analysis among nine representative ML techniques, identifying the ones that are more suited to VQA.

The remainder of this paper is organized as follows. Section~\ref{sec:database}, provides a state of the problem at hand, summarizing our earlier study of NR metrics. In section~\ref{sec:method}, the proposed predictive NR method is presented. The evaluation methodology is described in Section~\ref{sec:eval}. Our findings are discussed in Sections~\ref{sec:evalworst} to~\ref{sec:comptimes}, in relation to different test cases. The state-of-the-art on NR metrics in general and the use of ML techniques in particular, is given in Section~\ref{sec:background}. Finally, section~\ref{sec:conclusion} draws conclusions, highlighting our key contributions.

\section{Previous work}
\label{sec:database}

\begin{table}[t!]
\centering
\caption{Video dataset parameters range in terms of video types, compression and network packet loss ratio (960 samples in total).} 
 \begin{tabular}{|c|c|c|}
  \hline
  Video type&Compression&Packet loss\\
  \hline
  Blue Sky (bs1)&64kbps&PL$0\%$\\
  Mobile Calendar (mc1)&640kbps&PL$0.5\%$\\
  Pedestrian Area (pa1)&768kbps&PL$1\%$\\
  Park Run (pr1)&1024kbps&PL$1.5\%$\\
  River Bed (rb1)&2048kbps&PL$2\%$\\
  Sunflower (sf1)&3072kbps&PL$2.5\%$\\
  Shields (sh1)&4096kbps&PL$3\%$\\
  Station (st1)&5120kbps&PL$3.5\%$\\
  Tractor (tr1)&&PL$4\%$\\
  &&PL$4.5\%$\\
  &&PL$5\%$\\
  &&PL$10\%$\\
  \hline
 \end{tabular}
\label{tab:database}
\end{table}

\begin{table*}[t!]
\centering
\caption{PCC correlations to VQM of the eight NR metrics and SSIM. Cell colors give qualitative correlation levels: green (best), yellow (median), and red (worst).}
 \begin{tabular}{|c|c|c|c|c|c|c|c|c||c|}
 \hline
 Type&CX&MO&NM&NR&BM&BR&BL&JE&{\cellcolor{Turquoise}SSIM}\\
 \hline
 bs1&0.168312&0.01102&-0.487889&0.118084&{\cellcolor{Yellow}-0.013234}&-0.637055&{\cellcolor{OliveGreen}0.439352}&{\cellcolor{Maroon}-0.701313}&{\cellcolor{Turquoise}0.734887}\\
 mc1&{\cellcolor{OliveGreen}0.66322}&0.017724&-0.64416&0.538208&-0.065169&{\cellcolor{Maroon}-0.818207}&{\cellcolor{Yellow}0.084655}&0.36792&{\cellcolor{Turquoise}0.9035}\\
 pa1&0.29149&{\cellcolor{Maroon}-0.028091}&{\cellcolor{OliveGreen}0.645745}&{\cellcolor{Yellow}0.456716}&0.110438&0.464662&0.441793&0.05716&{\cellcolor{Turquoise}0.88293}\\
 pr1&0.303617&-0.164135&{\cellcolor{Maroon}-0.703943}&{\cellcolor{Yellow}-0.121612}&0.009715&-0.200413&0.496992&{\cellcolor{OliveGreen}0.6075}&{\cellcolor{Turquoise}0.687722}\\
 rb1&0.533103&{\cellcolor{OliveGreen}0.570433}&0.4320174&0.513867&0.546452&{\cellcolor{Yellow}0.44004}&0.200401&{\cellcolor{Maroon}-0.594002}&{\cellcolor{Turquoise}0.255487}\\
 rh1&{\cellcolor{OliveGreen}0.39129}&-0.474693&{\cellcolor{Yellow}0.165556}&0.320134&0.351533&0.369301&{\cellcolor{Maroon}-0.686387}&-0.670956&{\cellcolor{Turquoise}0.906884}\\
 sf1&0.412736&-0.414118&{\cellcolor{Maroon}-0.728497}&{\cellcolor{Yellow}0.13638}&0.516196&0.41971&{\cellcolor{OliveGreen}0.552045}&-0.41496&{\cellcolor{Turquoise}0.839916}\\
 sh1&0.412736&-0.09248&-0.351639&0.468456&{\cellcolor{Yellow}0.215916}&{\cellcolor{Maroon}-0.7191}&0.469199&{\cellcolor{OliveGreen}0.53221}&{\cellcolor{Turquoise}0.873088}\\
 st1&0.46892&-0.328997&{\cellcolor{Maroon}-0.646985}&0.350274&0.437243&{\cellcolor{Yellow}-0.208797}&{\cellcolor{OliveGreen}0.632241}&-0.267012&{\cellcolor{Turquoise}0.755406}\\
 tr1&0.5307191&{\cellcolor{Maroon}-0.17782}&0.087274&{\cellcolor{OliveGreen}0.737649}&0.511014&0.157158&{\cellcolor{Yellow}0.307524}&0.581363&{\cellcolor{Turquoise}0.885233}\\
\hline
Overall&{\cellcolor{OliveGreen}0.4176141}&-0.10811595&{\cellcolor{Maroon}-0.22325228}&0.35181569&{\cellcolor{Yellow}0.26201}&-0.07326658&0.29378164&-0.05020884&{\cellcolor{Turquoise}0.77250134}\\
&{\cellcolor{OliveGreen}$\pm$0.13449}&$\pm$0.27962&{\cellcolor{Maroon}$\pm$0.4859734}&$\pm$0.23678&{\cellcolor{Yellow}$\pm$0.227552}&$\pm$0.48674&$\pm$0.36191&$\pm$0.51401&{\cellcolor{Turquoise}$\pm$0.187342}\\
\hline
 \end{tabular}
  \label{tab:nrmetrics}
\end{table*}

The experimental survey we presented in~\cite{TorresES2016} served as motivation and starting point for this work. Our purpose was to study the performance of low complexity NR metrics in the assessment of network-impaired video quality and, if possible, to pinpoint NR features which could serve as alternative to FR metrics in situations with thin clients (such as mobile devices) or where real-time quality assessment is required. We studied eight well-known NR metrics, over a wide range of video types, compressions and lossy network conditions, benchmarking the NR assessments against the FR metric VQM. We concluded that none of the NR metrics was able to perform an accurate assessment on a general base, i.e. over all video types, compressions and network conditions. In that way, no NR metric could serve as alternative to the highly complex FR methods. Most importantly, all metrics failed under lossy networks. However, it also emerged, that each metric exhibited specific operational boundaries, within which the performance was accurate to the benchmark. Armed with these results, our next research hypothesis was that it would theoretically be possible to derive a hybrid NR metric characterized by a much broader operational boundary. Before we introduce this new metric (Section~\ref{sec:method}), it will help to summarize the key methodology and findings detailed in~\cite{TorresES2016}.

\begin{figure}[h!]
\begin{subfigure}{0.5\textwidth}
\includegraphics[width=\textwidth]{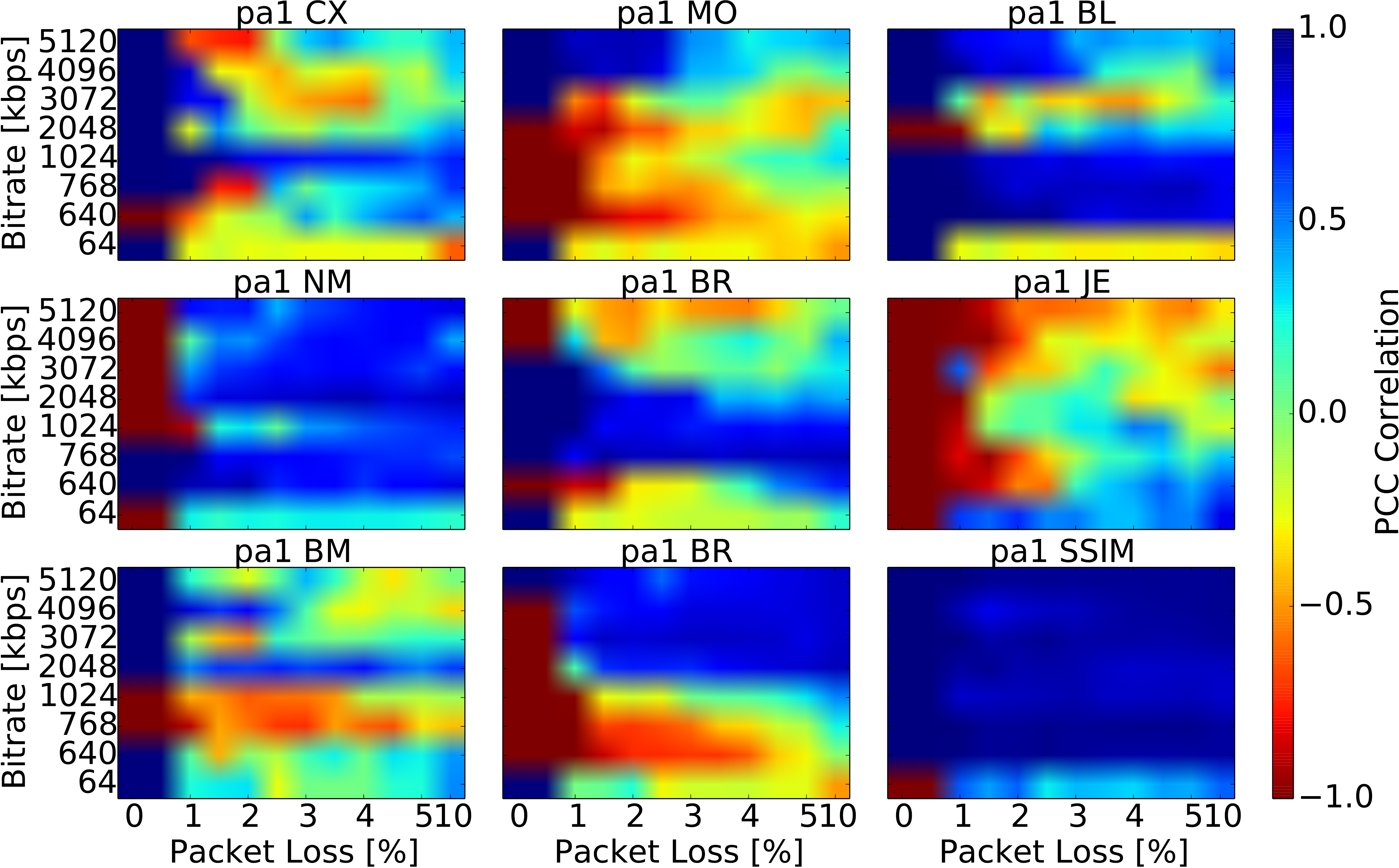}
\caption{Correlation maps for video pa1.}
\label{fig:corrpa1}
\end{subfigure}
\begin{subfigure}{0.5\textwidth}
\includegraphics[width=\textwidth]{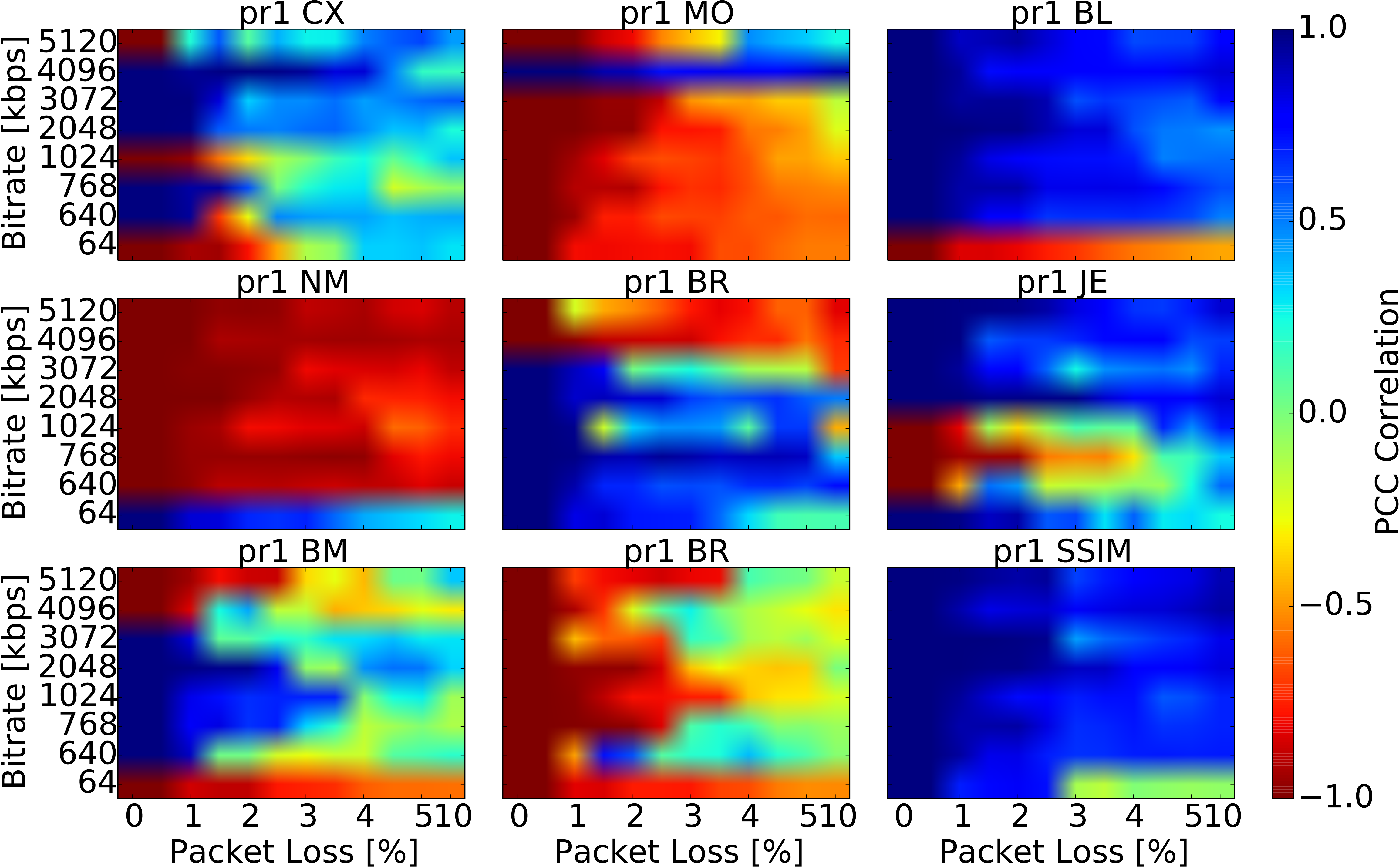}
\caption{Correlation maps for video pr1.}
\label{fig:corrpr1}
\end{subfigure}
\begin{subfigure}{0.5\textwidth}
\includegraphics[width=\textwidth]{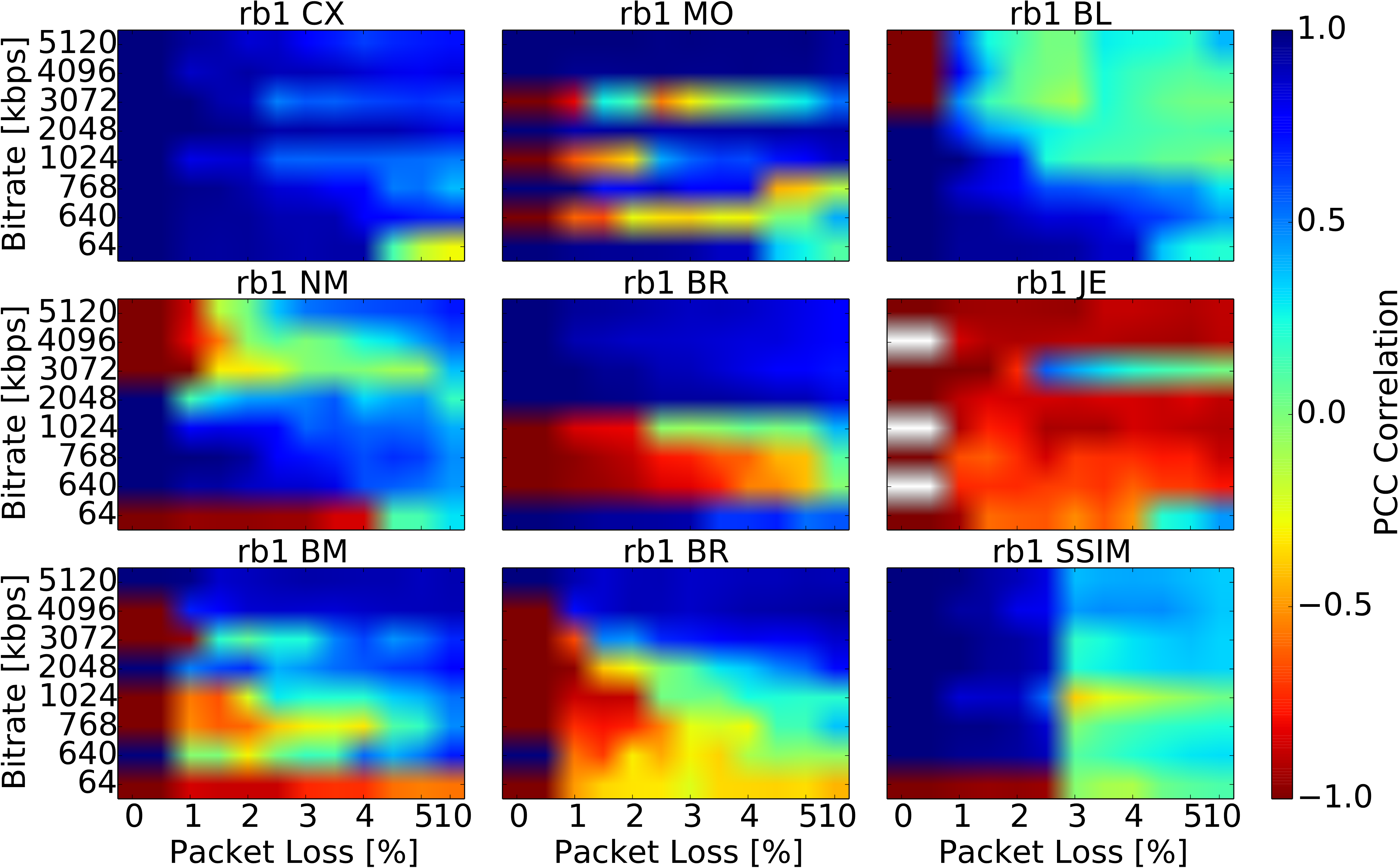}
\caption{Correlation maps for video rb1.}
\label{fig:corrrb1}
\end{subfigure}
\caption{Pearson correlation to VQM of the eight NR metrics (CX, MO, NM, NR, BM, BR, BL, JE) and the SSIM FR metric, considering bitrates between 64 and 5,120 Kbps and packet losses between 0 and 10\%. Video types: a) Pedestrian Area (pa1); b) Park run (pr1); and c) River bed (rb1). The original (unimpaired) videos were obtained from the Live Quality Video Database~\cite{Seshadrinathan:2010:SSO:1827592.1827596}. Network impairments were incurred by streaming videos through the PacketStorm network emulator~\cite{packetstorm}.\label{fig:nrmet}}   
\end{figure}

We studied eight NR features, namely complexity (CX), motion (MO), blockiness (BL), jerkiness (JE), average blur (BM), blur ratio (BR), average noise (NM) and noise ratio (NR). We also included SSIM, a well-known FR algorithm, which is less accurate and complex than the VQM benchmark~\cite{vqm}. All these metrics were evaluated over a range of $0-10\%$ packet loss rates, to account for one of the most critical types of network impairments~\cite{suarez2015}. The other parameters were video type and bitrate. The resulting $960$ samples of the video dataset (as detailed in Table~\ref{tab:database}) were correlated to VQM using the Pearson correlation index (PCC) ~\cite{Kendall:1987:KAT:59556}.

Ten original video types were obtained from the Live Quality Video Database~\cite{Seshadrinathan:2010:SSO:1827592.1827596}. We compressed (MPEG4/H.264) them at eight levels and impaired them at twelve packet loss rates, obtaining 960 videos.\footnote{Upon acceptance of this paper, we shall release the whole dataset and software implementation at www.tue.nl/universiteit/faculteiten/electrical-engineering/onderzoek/onderzoeksgroepen/electro-optical-communications-eco/research/network-management-and-control/datasets/network-impaired-video-dataset/.}  
 
The overall correlation to VQM (averaged across all dataset) of each of the eight NR metrics and SSIM is summarized in Table~\ref{tab:nrmetrics}. While rows one to ten of the table show the results for each of the specific video types, in row eleven the overall averaged and deviation correlation values can be seen. 

Looking at the overall correlations (last row), we observe that none of the NR metrics achieves an acceptable correlation ($50\%$). The best NR performant is the complexity (CX), with roughly an average correlation of 42\%. Blockiness (BL) and blur ratio (BR) reach roughly 30\%, while the average noise (NM) anti-correlates to the benchmark. Also, the standard deviations are noticeably high in all cases, which denotes a broad performance variation across the video dataset. This can be seen directly by looking at the spread of the cell values and colors.

As expected, being an FR metric, SSIM gives much better performance than any of the NR ones (rightmost column), with an overall correlation to VQM of about $77\%$. Yet the standard deviation is still relatively high, indicating that SSIM too will have a limited operational boundary. Further evaluations unveiled that in fact SSIM starts failing at high packet losses~\cite{TorresES2016}, depending on the video type.

In order to narrow down the working limits of the various NR metrics, in~\cite{TorresES2016} we went on analyzing the different video types individually (Figure~\ref{fig:nrmet}), with particular attention to compression level (Y axes) and packet loss (X axes). In Figure~\ref{fig:nrmet}, maximum correlation to VQM is shown in dark blue, while maximum anti-correlation is in dark red. Again we see that, although the analysis has been narrowed down (instead of being averaged across the whole dataset), none of the metrics operates accurately beyond some fairly narrow conditions. 

It is encouraging, though, that specific blue (well correlated) areas emerge. For example, in the pedestrian area video (pa1, Figure~\ref{fig:corrpa1}) blockiness performs well at low bitrates and on a broad range of packet loss. In park run (pr1, Figure~\ref{fig:corrpr1}), the noise ratio performs well on medium to low bitrates, but only when packet loss is low.  At the same time, jerkiness offers good complementary conditions (high bitrate, broad range of packet loss). 
These results encouraged us to pursue the study of hybrid metrics that would combine the strengths of individual metrics, as explained in the remainder.

\section{Predictive NR Video Quality Method}
\label{sec:method}
\begin{figure}[h!]
\centering
\includegraphics[width=0.5\textwidth]{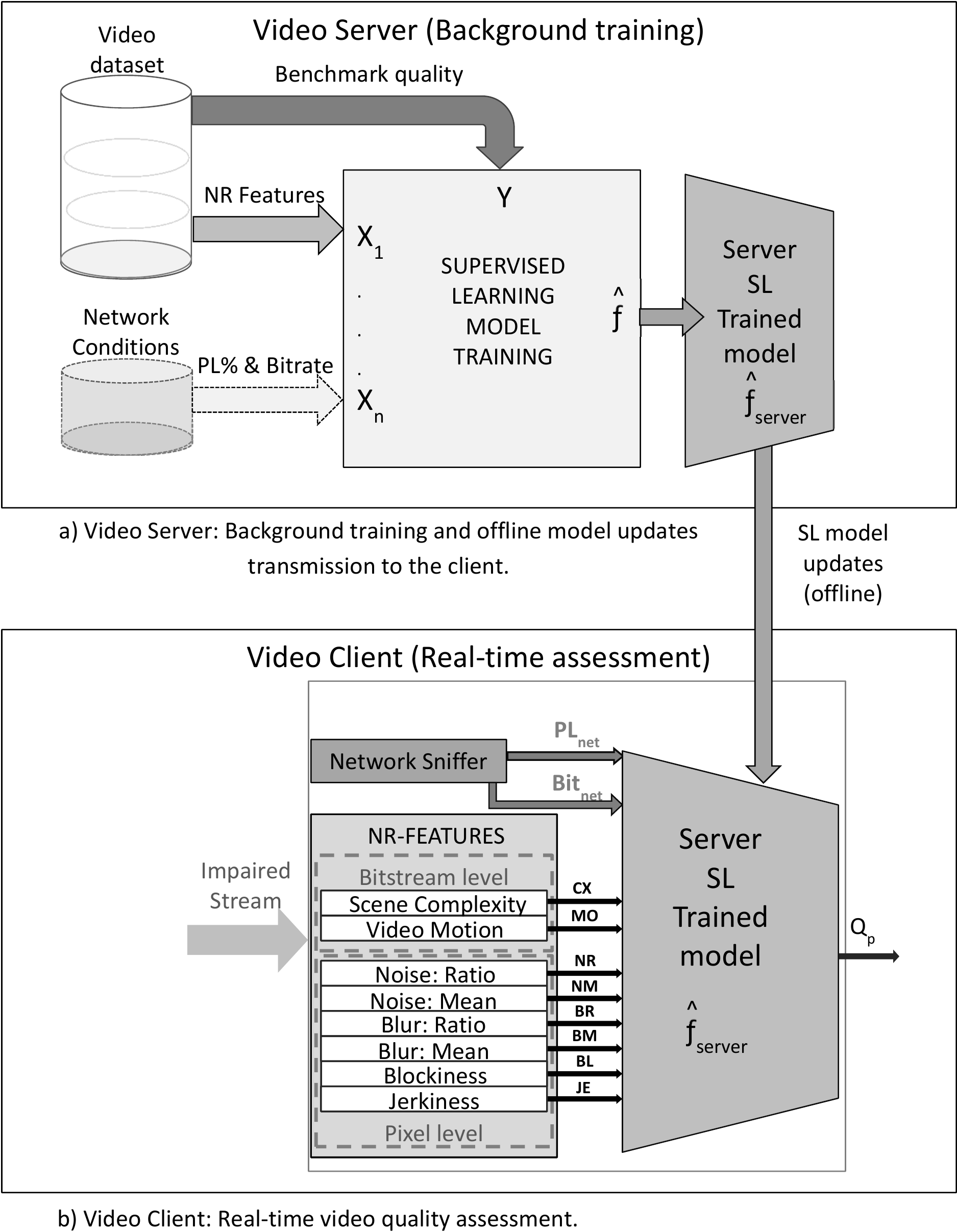}
\caption{\label{fig:serverclient}Block diagram of predictive the NR video quality assessment method. a) server side (background training); b) client side (real-time assessment).}
\end{figure}

In this section, we present our predictive NR video quality method. Figure~\ref{fig:serverclient} shows the block diagrams for the processes running, respectively, on the server side and in the clients. 

As with any prediction-based method, the accuracy of the model will substantially depend on the characteristics of the dataset used for training. In the case of our video service, the training set is composed by a number of video type samples stored in the server. Each sample in the training set includes the eight NR features of Table~\ref{tab:nrmetrics} (both in the pixel and the bitstream layers), two network condition parameters (packet loss rate and bitrate) and the ground truth quality index (without loss of generality, we used VQM in the present study, thanks to its proven correlation to subjective tests~\cite{journals/tbc/ChikkerurSRK11}). This training set is used (in the server) to maintain the quality prediction function, which is then employed on the client side to compute our predictive NR video quality assessment metric. 

At service launch, the service provider will already have a representative video types set (e.g., sport, action movies, cartoons, and so forth); thus an initial prediction model can be constructed (and made available to the client side). When a completely new video type is added, the prediction model will be less accurate. Yet, over the time the model will be updated based on new types and, what is more important, the chances of getting new video types will rapidly diminish. In this way, the server runs a process in the background in which the SL model is trained with the available video samples and new models ($\hat f_{server}$) are uploaded to the clients (on a continuous or periodic basis).

On the other end of the transmission link, the video client employs the SL model trained by the server, to generate its prediction-based quality metric ($Q_{p}$). During a streaming session, the client characterizes the incoming video in terms of NR features and real-time network conditions, matching this information against the prediction model to generate the quality estimation.    

Selecting the features that better characterize the video streams, are effective in the SL training process (in the server) and, ultimately, generate an accurate quality metric (in the clients), is not trivial. Our choice was driven by a preliminary (extensive) evaluation of classic NR metrics (Section~\ref{sec:database}~\cite{TorresES2016}), where we studied their operational boundaries. We followed the intuition that, if individual metrics would work accurately under specific conditions, a functional combination of those metrics could work on a broader range of conditions. Next, we hypothesized that machine learning could be a suitable method to extrapolate efficient NR quality assessment. 

The next step was to select representative features (of video stream and network conditions) for the purpose of training in a Supervised Learning (SL) manner. In general, a video stream can be characterized by several parameters, i.e. the ones that would allow differentiating among different video types. Parameters regarding the video scene composition have been demonstrated to  affect quality to a large extent~\cite{Liotta:2013:IVQ:2536853.2536903}. Among these, scene complexity and video motion have proven to correlate well with video quality~\cite{videotypes}. Scene complexity is defined as the number of objects or elements present in the frame, whereas video motion is the amount of movement in the video~\cite{videotypes}. Both features can be empirically obtained from the codec~\cite{Liotta:2013:IVQ:2536853.2536903}. On the pixel level, noise and blur components (mean and ratio per feature) have been demonstrated to provide a good measure of degradations in a frame-by-frame assessment~\cite{Choi_no-referenceimage}. In the same way, blockiness~\cite{PerraBl2014,GBIM}, described as a discontinuity between adjacent blocks in images and video frames~\cite{journals/spic/HemamiR10}, was demonstrated in our earlier study~\cite{TorresES2016} to show promising results. Finally, measuring the inter-frame degradations becomes fundamental in the presence of network impaired video. To this end, temporal features such as the Jerkiness (non-fluent and non-smooth presentation of frames) become fundamental~\cite{BorerJ2014}. Before they could be directly applied in the SL process, these eight NR metrics were averaged across the video and normalized between 0 and 1. Further details on how to compute these metrics are given in~\cite{TorresES2016}. 

In addition to the video stream characteristics, we chose two network features (packet loss and bitrate) to capture the most significant transmission effects on video quality~\cite{suarez2015}. These two parameters are calculated on the client side, during video reception, and are added to the other input features of the learning algorithm (i.e. the eight NR metrics).

These ten parameters conform the full characterization of the videos and serve both for training the SL model (offline on the server side) and for predicting the quality of the real-time received videos (in the client side, in the form of inputs to the trained SL method).  Through SL, we derive the quality prediction model (i.e. the function $\hat f_{server}$ in Figure~\ref{fig:serverclient}) by mapping input-output pairs of the training data. The model is then used to estimate the video quality, determining a suitable output value for any incoming stream (regardless of whether or not this has been part of the training set)~\cite{Mohri:2012:FML:2371238}. 

Our method, as described in Figure~\ref{fig:serverclient}, is generic and may be easily extended to explore different training features and benchmark quality (FR models or subjective studies), different video datasets and different SL algorithms. The details of our experimental evaluations, including the choice of the different SL algorithms are given next.

\section{Evaluation Methodology}\label{sec:eval}
We describe here the complete methodology used to evaluate the predictive quality metric introduced in Section~\ref{sec:method}. The experimental test-bed (Section~\ref{sec:testbed}, Figure~\ref{fig:evmethod}) comprises all the components used to carry out a comparative evaluation with the benchmark quality metric (VQM). The prediction model (i.e. the $\hat f_{server}$) is computed offline (as per Figure~\ref{fig:serverclient}a), exploring a whole range of machine learning options, as detailed in Section~\ref{sec:slalg}. Our method is generic, it does not demand a specific learning algorithm or benchmark algorithm. We have adopted VQM as our benchmark due to its demonstrated high correlation to the human vision system~\cite{journals/tbc/ChikkerurSRK11,vqm}. 

\subsection{Experimental Test-bed}\label{sec:testbed}
\begin{figure*}[t!]
 \centering
 \includegraphics[width=0.75\textwidth]{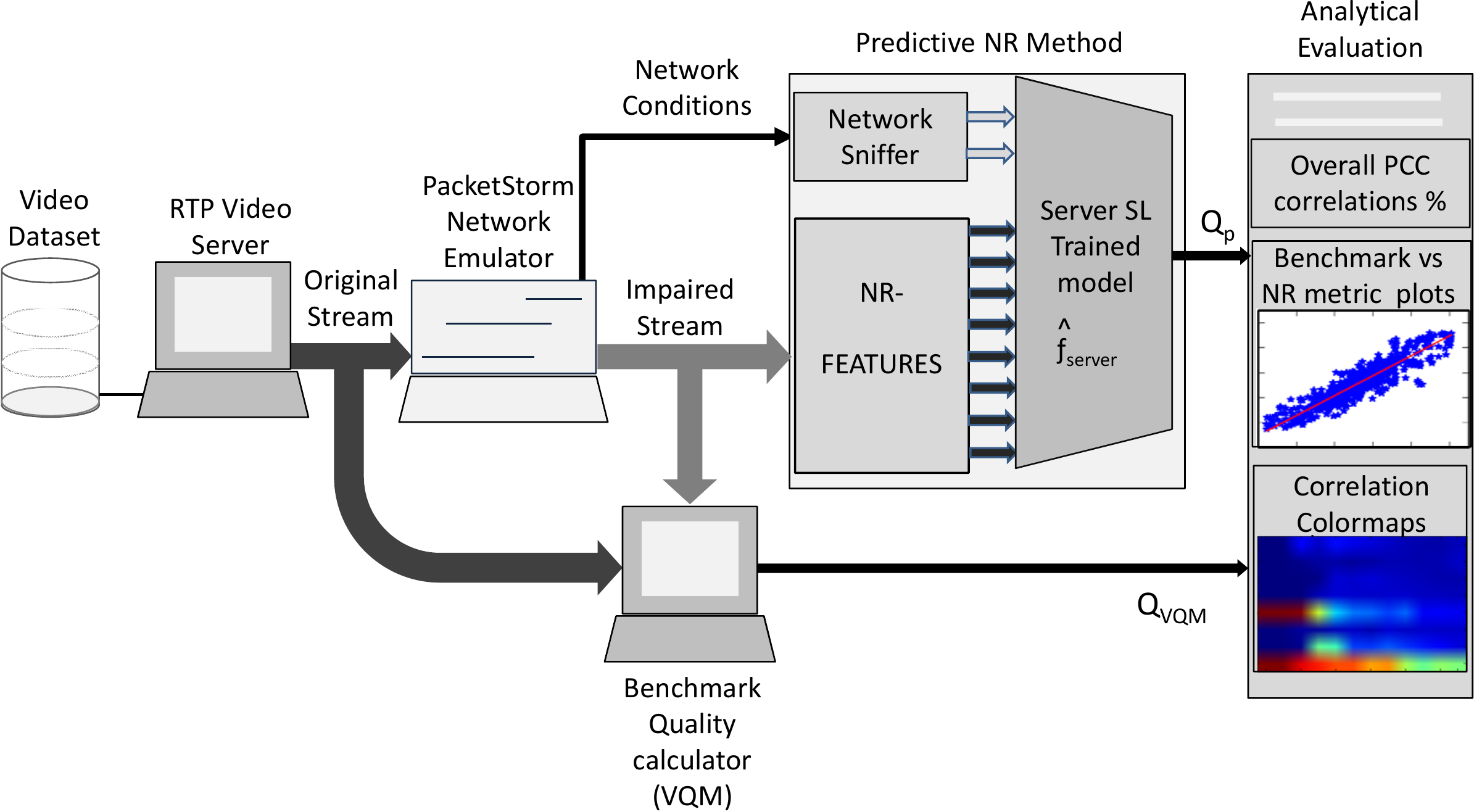}
 \caption{Evaluation test-bed.}
 \label{fig:evmethod}
\end{figure*}

Once the quality prediction function ($\hat f_{server}$) has been computed (offline), we are ready to perform real-time streaming tests, based on the components depicted in Figure~\ref{fig:evmethod}. Our testbed allows streaming any of the dataset videos on demand between the server and the client. We used an RTP video server to handle the streaming process, and a commercial network emulator (PacketStorm Hurricane II)~\footnote{http://packetstorm.com/packetstorm-products/hurricane-ii-software/} to shape and impair the stream in a controlled (replicable) environment.

The network-impaired stream is then fed to our client application, which generates the predicted metric $Q_p$. In parallel, we generate the benchmark quality index $Q_{vqm}$. We stream all videos, in turn, under a range of network conditions (Table~\ref{tab:database}), obtaining a full range of quality values, ready for statistical analysis. The accuracy is measured by means of a Pearson correlation (PCC)~\cite{Kendall:1987:KAT:59556} between the predicted quality and the benchmark quality.  

\subsection{Supervised learning methodology}\label{sec:slalg}


Given the broad variety of machine learning approaches in the literature, an important element of our work was to explore different algorithms and find suitable avenues. To this end, our experimental framework (Figure~\ref{fig:evmethod}) is sufficiently generic to perform tests on any type of SL algorithms (we have not included unsupervised learning methods in our study).

Among the well-established SL methods, we started experimenting with 16 different ones, ending up with a selection of nine methods that cover a representative set of algorithms, ranging from the least complex (towards the top of Table~\ref{tab:params}) to the most complex ones (towards the bottom of Table~\ref{tab:params}). Methods may be broadly categorized in two. Firstly, the white-box methods are able to capture a comprehensible relation between input and output features. Thus input-output connections are modelled in a straightforward way and can be interpreted by a human operator. On the other hand, black-box methods do not offer such relation and do not help understanding how certain predictions are derived. We review below the key features of the methods under scrutiny. 

One of the most known and simplest white boxes is linear regression~\cite{Free:2005}, which attempts to model the relationship between a scalar (output) and one or more independent variables  by means of a linear multidimensional model of the input data. 

Decision trees learning uses a decision tree as a predictive model which maps observations about an item to conclusions about the item's target value~\cite{id3}. They are classified according to the type of output provided. 

On the one hand, tree models, where the target variable takes a value from a finite set, are called classification trees. Leaves represent class labels and branches, conjunctions of features that lead to those class labels. On the other hand, decision trees, where the target variable can take continuous values (typically real numbers), are called regression trees.

The performance of regression and decision trees can be further improved by means of an ensemble approach. Ensembles use multiple learning algorithms to obtain better predictive performance than could be obtained from any of the constituent learning algorithms~\cite{journals/jair/OpitzM99}. 
Evaluating the prediction of an ensemble typically requires more computation than evaluating the prediction of a single model. Thus ensembles are mostly used as a way to compensate for poor learning algorithms by performing extra computation. For this reason, fast (less accurate) algorithms such as decision trees are commonly used with ensembles. 

Since the first conception, several approaches to combine the ML models have appeared. One early method is the Bootstrap aggregating~\cite{Breiman:1996:BP:231986.231989}, often abbreviated as bagging, which involves having each model in the ensemble vote with equal weight. Another method, Boosting~\cite{Breiman:96:TR}, involves incrementally building an ensemble by training each new model instance to emphasize the training instances that previous models misclassified. In some cases, boosting has been shown to yield better accuracy than bagging, but it also tends to be more likely to over-fit the training data. 

The most common implementation of Boosting is Adaboost~\cite{Freund:1997:DGO:261540.261549}. In Adaboost, short for "Adaptive Boosting", the output of the other learning algorithms ('weak learners') is combined into a weighted sum that represents the final output of the boosted classifier. 
While specific learning algorithms will tend to suit some particular problem types better than others, and will typically have many different parameters and configurations to be adjusted before achieving optimal performance on a dataset, AdaBoost (with decision trees as the weak learners) is often referred to as the best out-of-the-box classifier. AdaBoost is used only for classification and thus in order to use it, the quality index range (0 to 1) needs to be converted into a finite set of values.

Another type of boosting known to work very well together with regression trees is LS-Boost (least squares)~\cite{friedman2000greedy}. Like other boosting methods, LS-boosting combines weak learners into a single strong learner, in an iterative fashion, where the goal is to learn the model that predicts the outputs while minimizing the mean squared error to the true values (averaged over the training set).

White boxes are appreciated for their comprehensive models. Yet, they have also been demonstrated to have limited predictive capacity or to be inflexible and computationally cumbersome. The best classification and regression accuracy is typically achieved by black-box models such as Gaussian processes or neural networks, or complicated ensembles of them~\cite{TurnerBB2015}. These models do not, in general, provide a clear explanation of the reasons as to how they have come to a certain prediction. 

The Gaussian Process Regression (or Kriging)~\cite{citeulike:3683223} provides an example. The basic idea of Kriging is to predict values by means of interpolation in which the interpolated values are modeled by a Gaussian process governed by prior covariances Under suitable assumptions on the priors, Kriging gives the best linear unbiased prediction of the intermediate values. 

Support Vector Machines (SVMs)~\cite{Cortes:1995:SN:218919.218929}) are supervised learning models with associated learning algorithms that analyze data used for classification and regression analysis. Given a set of training samples, each marked as belonging to one of two categories, an SVM training algorithm builds a model that assigns new samples into one category or the other, making it a non-probabilistic binary linear classifier. An SVM model is a representation of the samples as points in space, mapped so that the examples of the separate categories are divided by a clear gap that is as wide as possible. New samples are then mapped into that same space and predicted to belong to a category based on which side of the gap they fall in.

Finally, we tested artificial neural networks (ANNs)~\cite{Haykin:1998:NNC:521706}, a family of models inspired to biological neural networks, used to estimate or approximate functions that can depend on a large number of generally unknown inputs. ANNs are generally presented as systems of interconnected "neurons" which exchange messages between each other. The connections have numeric weights that can be tuned based on various optimization methods, making neural nets adaptive to inputs and capable of learning. The feedforward neural network was the first and simplest type of artificial neural network devised. In this case, the information moves in only one direction, forward, from the input nodes, through the hidden nodes (if any) and to the output nodes. A variation on the feedforward network is the cascade forward network which has additional connections from the input to every layer, and from each layer to all following layers.

For the purposes of our comparative analysis among the different SL techniques, we considered an increasing range of complexities, as shown in Table~\ref{tab:params} where the methods are ordered by complexity. We implemented these methods based on the ML toolbox~\cite{MLMatlab} and the Neural Network toolbox~\cite{ANNMatlab} of Matlab, and the library LIBSVM~\cite{CC01a} for the support vector regression model. 

\begin{table}[t!]
\centering
\caption{Parameters used for the different machine learning techniques}
 \begin{tabular}{|c|c|c|c|}
 \hline
 Type&Technique&Acronym&Parameters\\
 \hline
 &Multiple&&\\
 &Linear&LR&Added bias\\
 W&Regression&&\\
 \cline{2-4}
 H&Standard&&type:binary\\
 I&Regression&RT&N. Branches$>$15\\
 T&Tree&&\\
\cline{2-4}
 E&Ensemble&&N. Models:500\\
 &Regression&ERT-LSB&N. Branches$>$15\\
 &Tree LS-Boost&&Learning Rate:0.01\\
 \cline{2-4}
 B&Ensemble&&N. Models:500\\
 O&Regression&ERT-BR&N. Branches$>$15\\
 X&Tree Bagging&&\\ 
 \cline{2-4}
 &Ensemble&&N. Classes: 100 (1/100)\\
 &Decision&EDT-AB&N. Models:200\\
 &Tree Adaboost&&N. Branches$>$10\\
 &&&Learning Rate: 0.2\\
 \hline
 &Gaussian&&Method: exact\\
 B&Process&GPR&Basis: constant\\
 L&Regression&&Kernel: squaredexponential\\
 \cline{2-4}
 A&Support Vector&SVR&type: epsilon\\
 C&Regression&&kernel: radial basis\\
 K&&&cost: 20\\
 &&&epsilon:0.1\\
 \cline{2-4}
 B&FeedForward&FNN&N. Hidden Neurons: 20\\
 O&Neural Net.&&Training: Levenberg-Marquardt\\
\cline{2-4}
 X&Cascaded FW.&CNN&N. Hidden Neurons: 20\\
 &Neural Network&&Training: Levenberg-Marquardt\\
\hline
 \end{tabular}
  \label{tab:params}
\end{table}

Each algorithm requires the tuning of certain parameters in order to optimize their performance. The values included in Table~\ref{tab:params} (third column), have been found to perform better with our dataset. In order to perform the Multiple Linear Regression, we added a bias vector (a vector of all ones) to the input data. As we explained in the previous section, to use the ensemble decision tree with Adaboost, the dataset outputs have to be converted to a set of finite values. After careful experimentation, we set the number of classification classes to $100$, ranging for $0.00$ to $0.99$. Values are then rounded to their second decimal. 



Another important choice in performing machine learning experiments consists on the way the training set is picked out of the whole dataset. The method used is bound to have a sensitive effect on the performance of the prediction models and, ultimately, on the accuracy of the NR metric. To mimic typical situations faced by a video service provider, we carried out two set of experiments. Blind prediction, represents the worst-case performing scenario, whereby the video under consideration is unknown to the machine learning model (Section~\ref{sec:evalworst}). We also consider the performance of the more typical cases using random cross-validation tests (Section~\ref{sec:evalcom}). Finally, we studied the sensitivity of our metric to the size of the training set (Section~\ref{sec:sizetraining}).

More exactly, when a new video is made available in a video server, it is possible that the server administrator does not yet have video traces of it and thus cannot or does not want to re-train the ML model. In this case, when a client requests the new video, its quality will be blindly assessed against a model which does not include traces of any variant of it. In this set of experiments we wanted to assess the NR metric and in it the ML models performance in this worst case scenario. This case is covered in Section~\ref{sec:evalworst}.

In the most common scenario, the video server is able to prepare the ML model from samples of the whole data-set before being transmitted to the client  (Section~\ref{sec:evalcom}). In this way, characterizations of all the videos are present in the model of the system. In this second set of tests we put the ML models to test, to select the best overall performer. Furthermore, we decided to explore the dependence of the metric performance to the size of the training set (Section~\ref{sec:sizetraining}).

\section{Evaluation of the worst-case scenario: unknown video class, blind prediction}
\label{sec:evalworst}

\begin{table*}[t!]
\centering
\caption{Overall performance of nine machine learning algorithms in blind mode (worse-case scenario, 10-fold cross-validation). Values indicate PCC correlations to VQM, averaged for each video type across all compression levels and network conditions (96 cases). Cell colors give qualitative correlation levels: green (best); orange (median); and red (worst).}
 \begin{tabular}{|c|c|c|c|c|c|c|c|c|c|}
 \hline
 Type&LR&RT&ERT-LSB&ERT-BR&EDT-AB&GPR&SVR&FNN&CNN\\
 \hline
 bs1&0.812965&{\cellcolor{Yellow}0.85936}&{\cellcolor{OliveGreen}0.956393}&0.955472&0.736283&0.831816&{\cellcolor{Maroon}0.563584}&0.941159&0.955876\\
 mc1&{\cellcolor{Yellow}0.883728}&0.842941&{\cellcolor{OliveGreen}0.927742}&0.895166&0.584229&0.8668&{\cellcolor{Maroon}0.492795}&0.885119&0.919813\\
 pa1&0.871275&0.95522&{\cellcolor{OliveGreen}0.970599}&{\cellcolor{Yellow}0.916031}&0.819921&0.934233&{\cellcolor{Maroon}0.708535}&0.921276&0.954169\\
 pr1&0.89029&{\cellcolor{Yellow}0.690704}&0.768428&0.818467&0.285759&{\cellcolor{OliveGreen}0.868404}&0.243897&{\cellcolor{Maroon}-0.618838}&-0.288696\\
 rb1&0.886588&0.706741&0.901205&{\cellcolor{OliveGreen}0.949458}&0.506328&{\cellcolor{Yellow}0.808595}&0.705418&{\cellcolor{Maroon}0.447927}&0.868012\\
 rh1&-0.315646&0.78308&{\cellcolor{OliveGreen}0.797164}&0.771964&{\cellcolor{Maroon}-0.356914}&-0.117229&0.725025&{\cellcolor{Yellow}-0.029357}&-0.27582\\
 sf1&{\cellcolor{Yellow}0.943491}&0.929498&0.97286&{\cellcolor{OliveGreen}0.974271}&0.761469&0.954198&{\cellcolor{Maroon}0.746171}&0.915238&0.953164\\
 sh1&{\cellcolor{Yellow}0.848927}&0.828483&0.920604&{\cellcolor{OliveGreen}0.929359}&{\cellcolor{Maroon}0.616136}&0.926687&0.758201&0.671181&0.866666\\
 st1&{\cellcolor{Yellow}0.935718}&0.858198&{\cellcolor{OliveGreen}0.970538}&0.966459&0.708189&0.963359&{\cellcolor{Maroon}0.46614}&0.967263&0.820368\\
 tr1&0.923678&0.859136&0.961394&0.960631&{\cellcolor{Maroon}0.508367}&{\cellcolor{OliveGreen}0.975721}&0.711093&{\cellcolor{Yellow}0.937621}&0.944151\\
\hline
Overall&0.768101&0.830436&{\cellcolor{OliveGreen}0.914693}&0.913728&{\cellcolor{Maroon}0.516977}&{\cellcolor{Yellow}0.801259}&0.612086&0.603859&0.671770\\
&$\pm$0.382840&$\pm$0.084724&{\cellcolor{OliveGreen}$\pm$0.073893}&$\pm$0.067849&{\cellcolor{Maroon}$\pm$0.344464}&{\cellcolor{Yellow}$\pm$0.327775}&$\pm$0.167778&$\pm$0.533213&$\pm$0.504905\\
\hline
 \end{tabular}
  \label{tab:NR1video}
\end{table*}

The accuracy of the prediction model will substantially depend upon the characteristics of the dataset used for training. In our case, a set of video samples stored in the server, which will be used by the service provider to keep an up-to-date prediction function and, in turn, ensure that the predictive NR function stays accurate. As mentioned in Section~\ref{sec:method}, on service launch, the service provider will have a representative video types set and thus an initial model can be constructed and sent to the client. When, due to a completely new type of video, the prediction model is to be updated, the server will notify the client and the model in the client will be upgraded.

Therefore, the most typical scenario will see an up-to-date prediction model. This case will be evaluated in Section~\ref{sec:evalcom}. We now consider the worse-case scenario, to evaluate the bottom-line performance of our metric. To test SL in blind mode, the model is trained with nine (out of ten) video types and is tested on the 96 samples of the remaining one (8 compression levels and 12 network conditions). For statistical significance, we performed a 10-fold cross-validation test, evaluating in turn, each of the ten videos as a new (unknown) class.

The overall performance of the nine different machine learning algorithms in blind mode is detailed in Table~\ref{tab:NR1video}. The first striking result is that our metric always performs considerably better than any of the conventional NR metrics (Table~\ref{tab:nrmetrics}). The worst-case performance of the worst-performing machine learning algorithms (51.7\% EDT-AB Table~\ref{tab:NR1video}) was better than the best-performing NR metric (41.7\% CX Table~\ref{tab:nrmetrics}). The Ensemble Regression Tress methods achieve the best average performance of 91.3\% (ERT-BR) and 91.4\% (ERT-LSB).

Comparing the different machine learning algorithms, we found another important result: the white-box approaches (LR, RT, ERT-BR, ERT-LSB and EDT-AB) outperform the black-box ones (GPR, SVR, FNN and CNN). This is interesting because the former methods tend to be less computationally intensive. Intuitively, we can explain this result by looking at the standard deviations, which tend to be rather large (up to 53\% in FNN). This is to be expected in blind prediction when the samples are significantly different. In fact, the most distinctive videos (the ones with distinctive time and space complexity) were predicted with lower accuracy. For instance, pa1 is well-represented by the other nine video types: thus the 10-fold validation for pa1 leads to consistently accurate predictions (71 to 93\%). At the other end of the spectrum is video type pr1, which leads to diverse prediction accuracies (-62\% to 87\%). We must stress that these high variations are typical of blind prediction and will not appear in the most common operational condition (Section~\ref{sec:evalcom}). 

\begin{figure}[h!]
 \centering
 \includegraphics[width=0.5\textwidth]{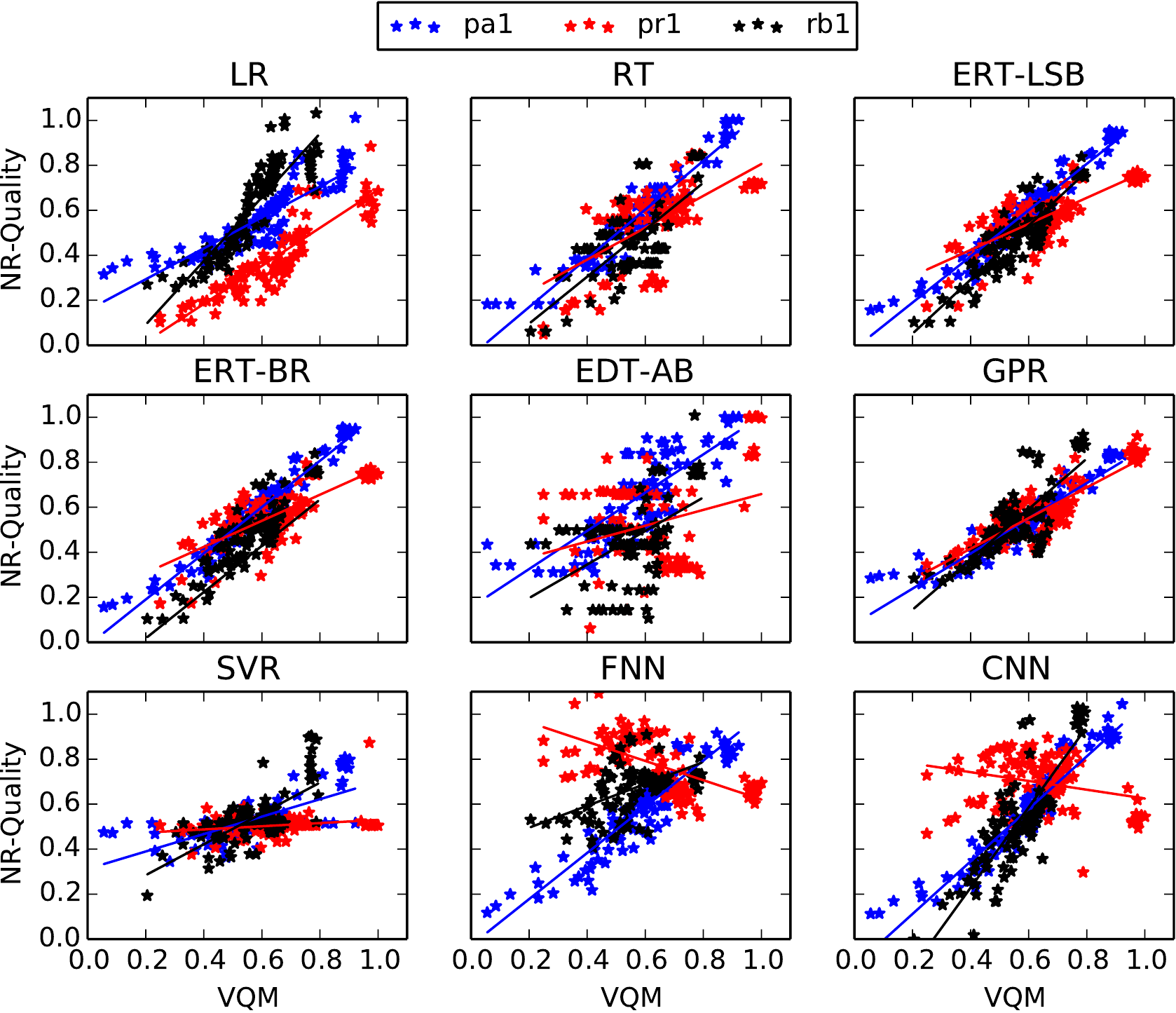}
 \caption{Correlation diagrams of nine different prediction algorithms (LR, RT, ERT-LSB, ERT-BR, EDT-AB, GPR, SVR, FNN, CNN) in comparison to VQM (used as benchmark). The three sample videos are: pa1 (blue stars); rb1 (black stars); and pr1 (red stars).}
 \label{fig:vqmvsm13videos}
\end{figure}

\begin{figure}[h!]
\begin{subfigure}{0.5\textwidth}
\includegraphics[width=\textwidth]{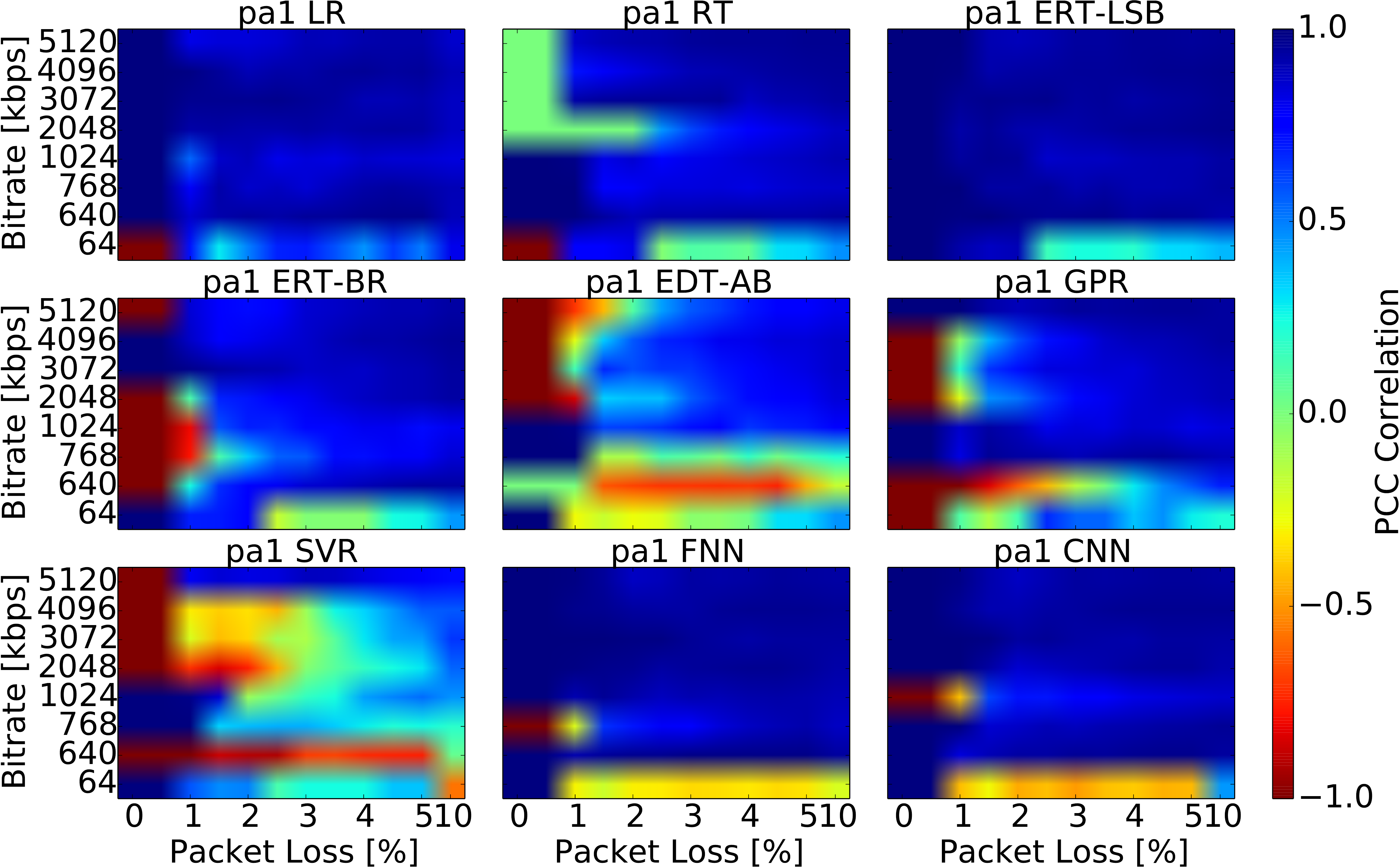}
\caption{Correlation maps for video pa1.}
\label{fig:nrselvideospa1}
\end{subfigure}
\begin{subfigure}{0.5\textwidth}
\includegraphics[width=\textwidth]{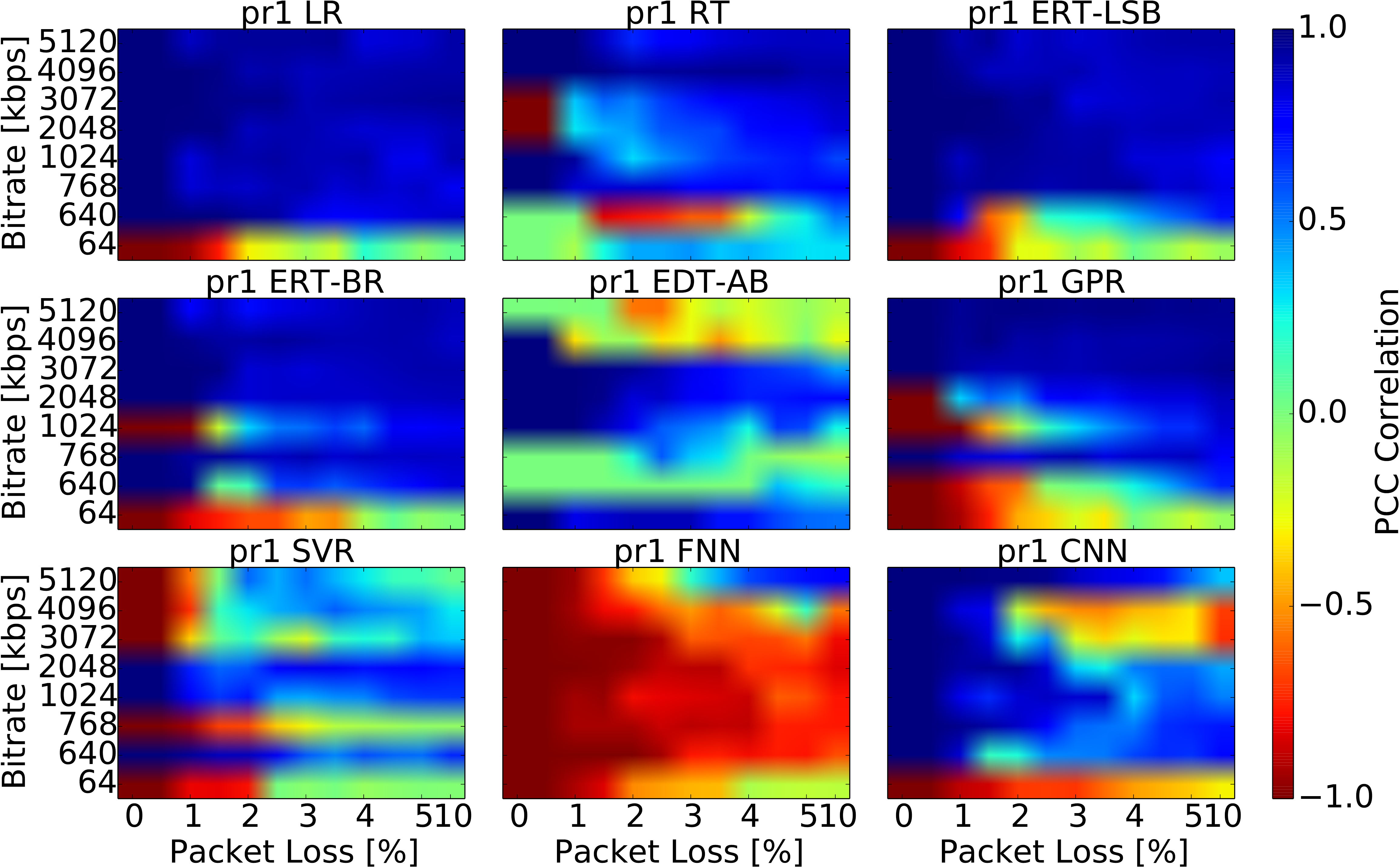}
\caption{Correlation maps for video pr1.}
\label{fig:nrselvideospr1}
\end{subfigure}
\begin{subfigure}{0.5\textwidth}
\includegraphics[width=\textwidth]{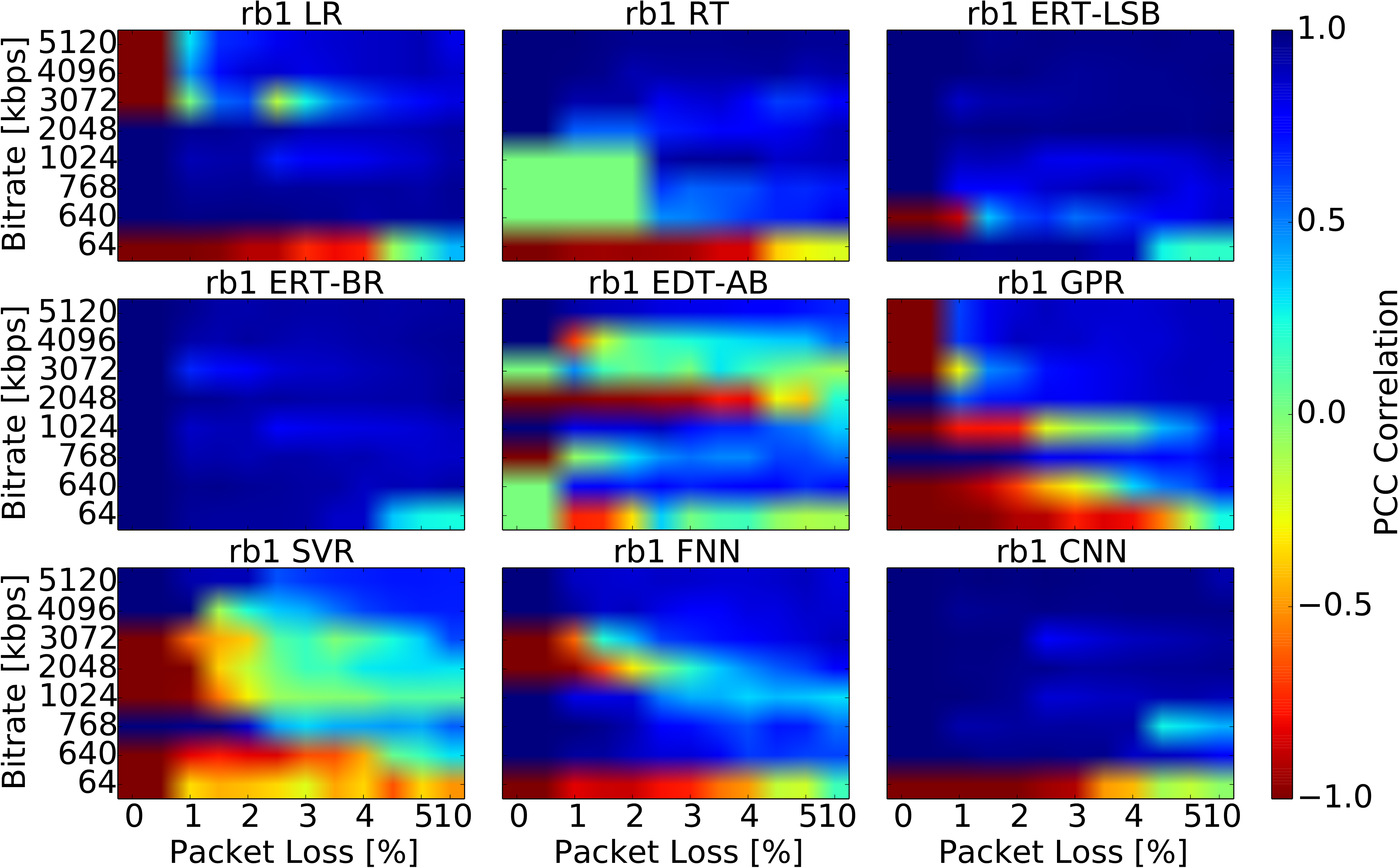}
\caption{Correlation maps for video rb1.}
\label{fig:nrselvideosrb1}
\end{subfigure}
\caption{Pearson correlation to VQM of the nine prediction algorithms (LR, RT, ERT-LSB,ERT-BR, EDT-AB, GPR, SVR, FNN, CNN), considering bitrates between 64 and 5,120 Kbps and packet losses between 0 and 10\%. Video types: a) Pedestrian Area (pa1); b) Park run (pr1); and c) River bed (rb1).\label{fig:corrml}}
\end{figure}

To better explore the differences across the test videos, Figure~\ref{fig:vqmvsm13videos} shows the correlation diagrams of the three most distinctive videos (pa1, pr1 and rb1), whose NR metrics were scrutinized in Section~\ref{sec:database} (Figure~\ref{fig:nrmet}). Each diagram picks one machine learning algorithm in relation to the benchmark VQM, showing the three video types in different colors. In this way, the most accurate predictions are concentrated around the main diagonal (y=x). We observe how video type pa1 (blue stars) is predicted consistently well, followed by rb1 (black stars). On the other hand, pr1 (red stars) is the most difficult to predict. Overall, RT and ERT-LB are the ones that deal the best with blind prediction; and in general, black box approaches perform the worst. Of these, only GPR performs consistently well on all videos, while the two neural networks (FNN and CNN) struggle with rb1 and fail with pr1. The support vector machine fails on all cases.

Finally, to better visualize the working range of the different machine learning algorithms, Figure~\ref{fig:corrml} shows the Pearson Correlation (PCC) colormaps analogous to those of Figure~\ref{fig:nrmet} (NR metrics). Strikingly, the well-correlated range (dark blue) extends much further (both in packet loss and bitrate levels) than the original NR metrics. The color patterns show also how the less complex machine learning methods (the upper maps in Figures~\ref{fig:nrselvideospa1},~\ref{fig:nrselvideospr1} and~\ref{fig:nrselvideosrb1}) have a broader operational range than the more complex algorithms (lower maps in Figure~\ref{fig:nrselvideospa1},~\ref{fig:nrselvideospr1} and~\ref{fig:nrselvideosrb1}). As we already hinted, the best performers are the Ensemble Regression Trees, particularly LS-Boost (ERT-LSB) achieves nearly full correlation for all bitrates and network conditions.

\begin{figure}[h!]
 \centering
 \includegraphics[width=0.445\textwidth]{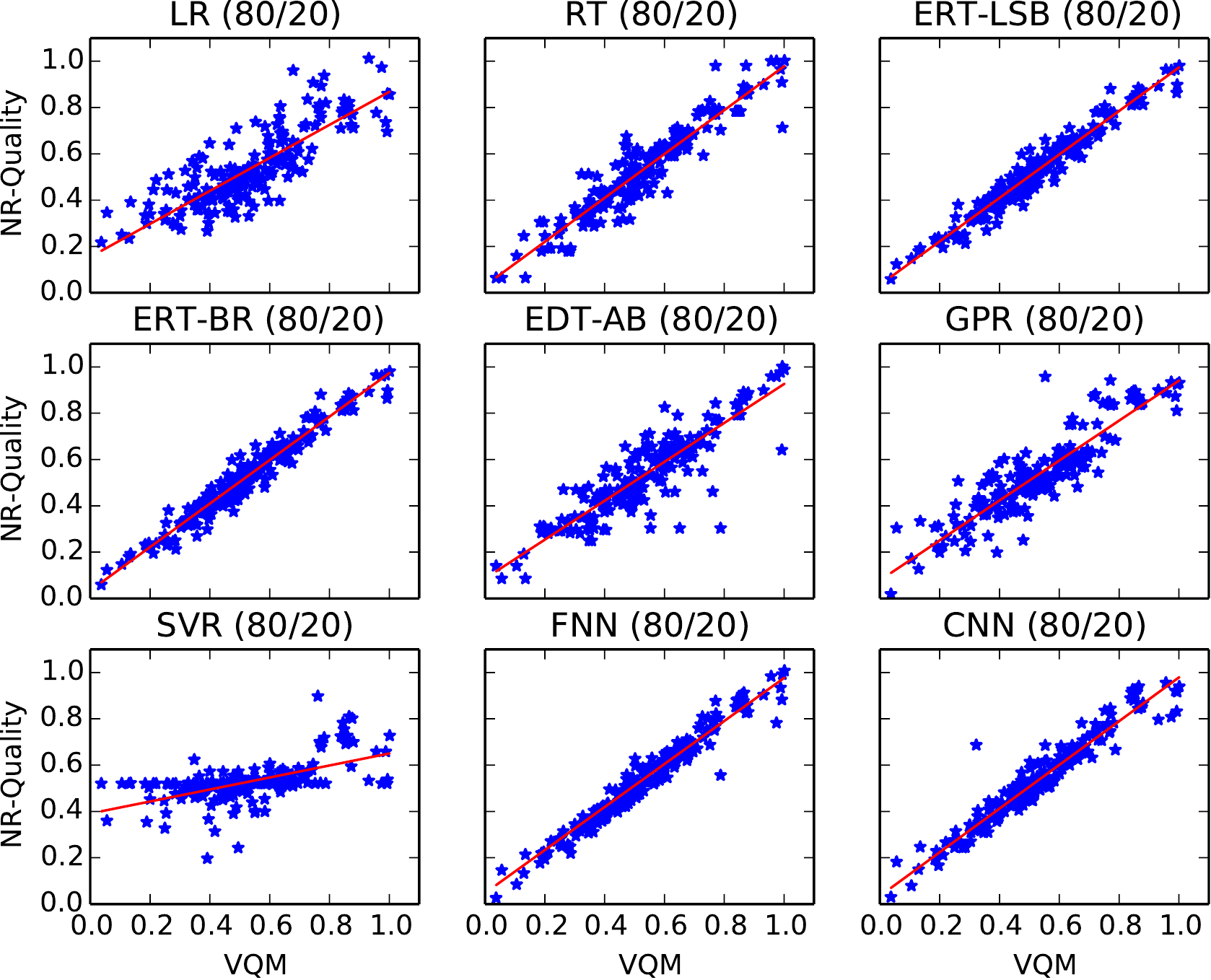}
 \caption{80\%-20\% training to testing data distribution. The diagrams show the overall correlation diagrams of nine different prediction algorithms (LR, RT, ERT-LSB, ERT-BR, EDT-AB, GPR, SVR, FNN, CNN) in comparison to VQM (used as benchmark).}
 \label{fig:randomregval8020}
\end{figure}

\begin{figure}[h!]
\begin{subfigure}{0.445\textwidth}
 \includegraphics[width=\textwidth]{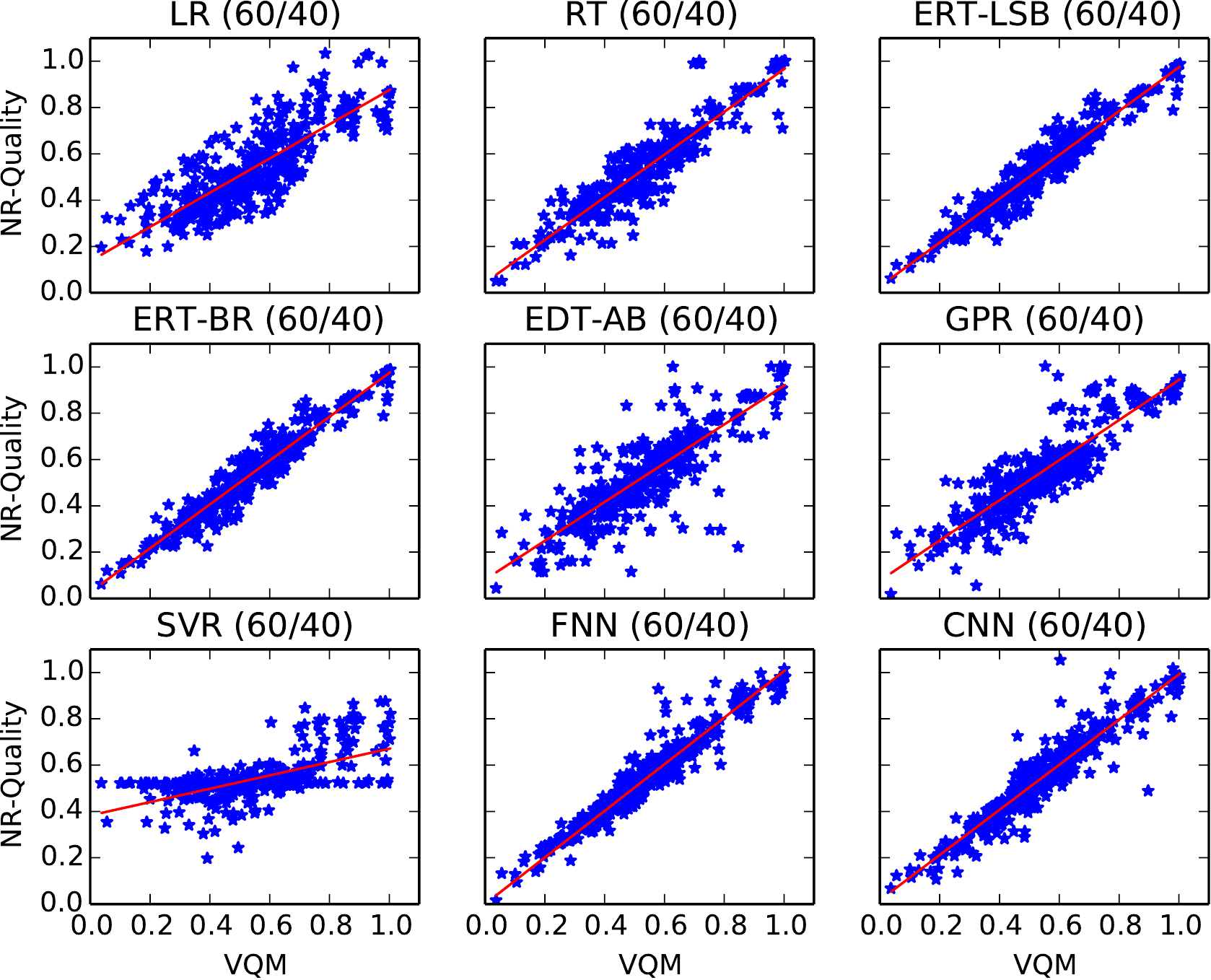}
 \caption{Data distribution: 60\%-40\%.}
 \label{fig:randomregval6040}
\end{subfigure}
\begin{subfigure}{0.445\textwidth}
 \includegraphics[width=\textwidth]{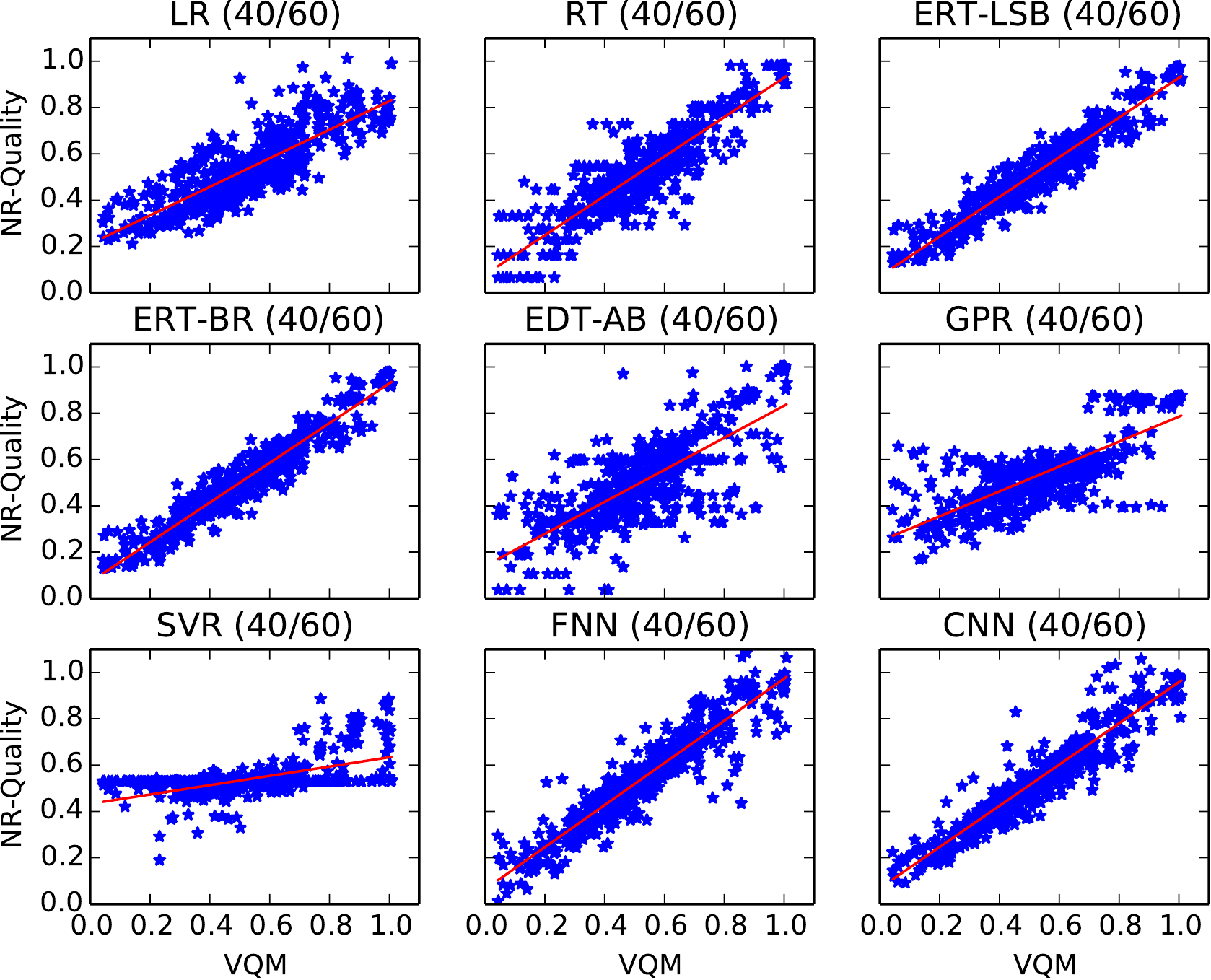}
 \caption{Data distribution: 40\%-60\%.}
 \label{fig:randomregval4060}
\end{subfigure}
\begin{subfigure}{0.445\textwidth}
 \includegraphics[width=\textwidth]{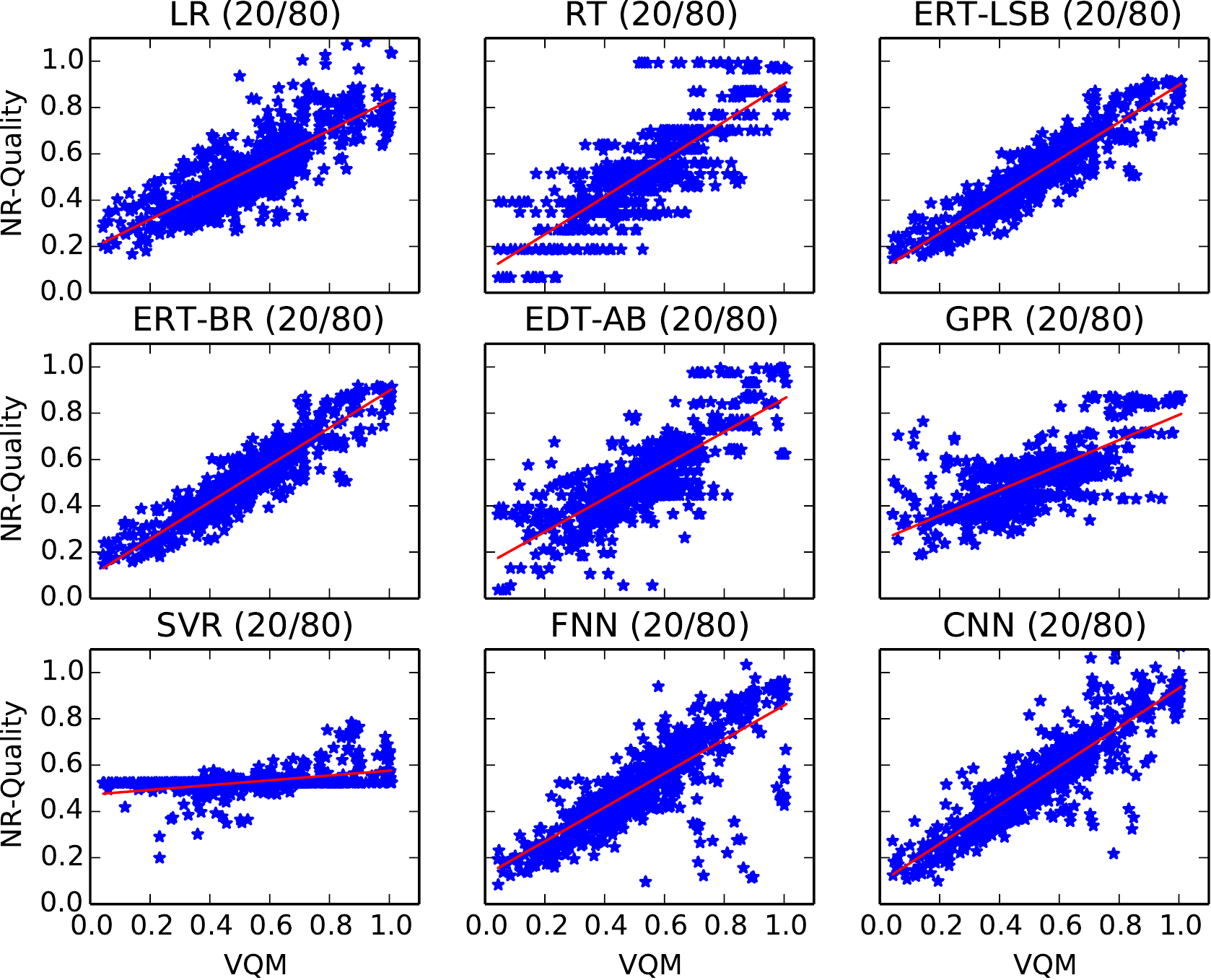}
 \caption{Data distribution: 20\%-80\%.}
 \label{fig:randomregval2080}
\end{subfigure}
\caption{Predicted values vs benchmark quality (VQM) for different distributions of training and testing data: a) 60\%-40\%; b) 40\%-60\%; c) 20\%-80\%.}
\label{fig:sizes}
\end{figure}

\begin{figure}[h!]
 \centering
  \includegraphics[width=0.49\textwidth]{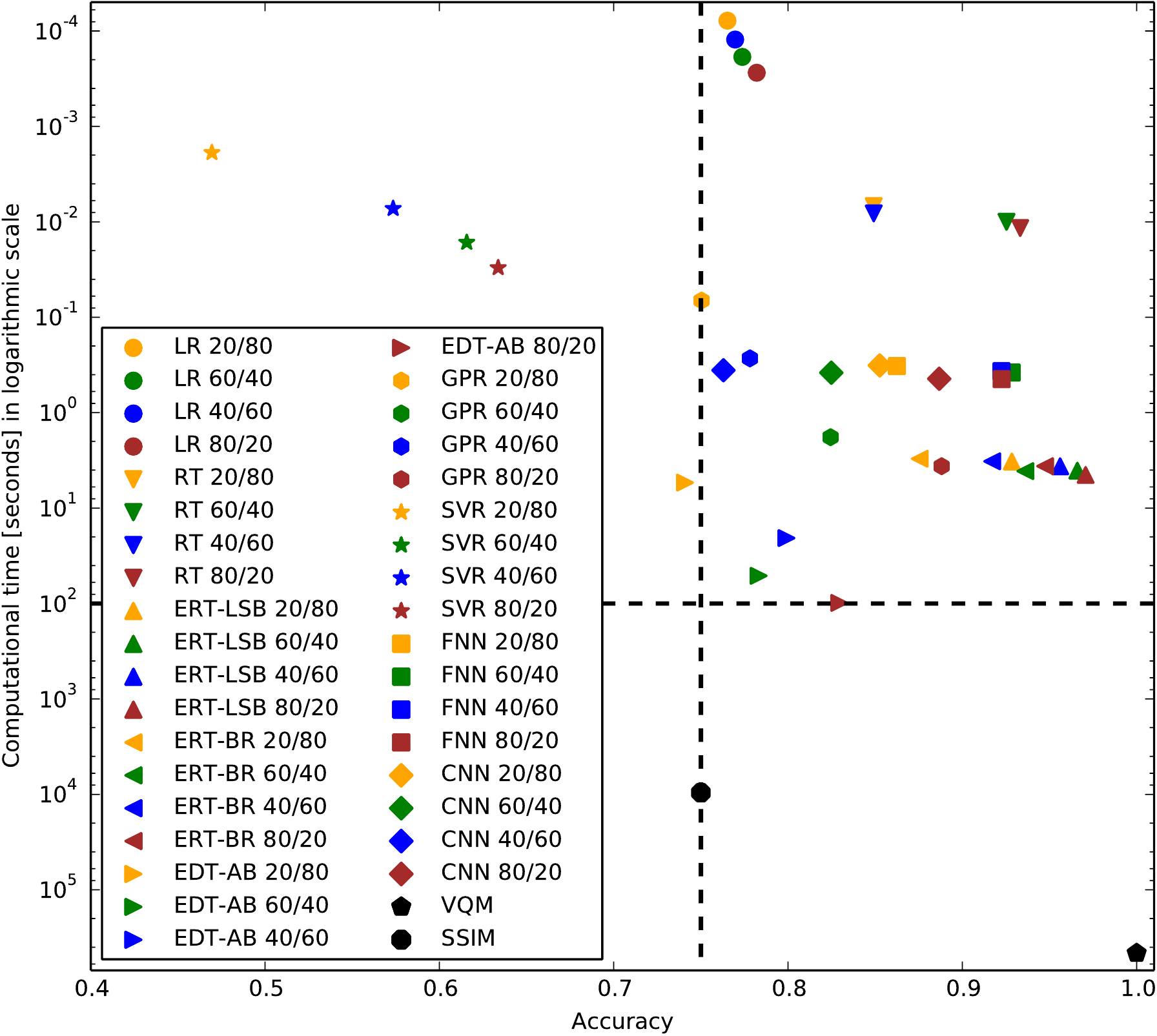}
 \caption{Performance trade-offs (accuracy and computational time) of the different learning algorithms (colored symbols), considering all training/testing combinations. SSIM and VQM (in black) are used to benchmark our metrics. With the exception of SVR, all other learning models perform much better than SSIM. Computational time (in log scale) is 4 orders smaller than SSIM and 6 orders smaller than VQM.}
 \label{fig:timesbest}
\end{figure}

\section{Evaluation of common-case scenario: known video class, prediction based on prior video traces}
\label{sec:evalcom}
We evaluate here the typical scenario in which our prediction based metric is assessed on video conditions (type, rate and packet loss level) that have previously been seen by the SL algorithm. Thus, we can assume that the prediction model will have been trained on samples from all the videos belonging to the service provider's dataset. Our aim is to evaluate the performance of our metric (as described in Section~\ref{sec:method}) under realistic conditions, for a representative set of machine learning algorithms (LR, RT, ERT-LSB, ERT-BR, EDT-ADT, GPR, SVR, FNN, CNN).

We follow a standard machine learning evaluation method. We randomize the whole dataset (960 samples), splitting it into five subsets (192 samples each). On each of the nine machine learning algorithms, we perform a 5-fold cross-validation test, using in turn one subset for testing and the other four for training. Just like in blind prediction (Section~\ref{sec:evalworst}), the resulting nine prediction models are used to find Pearson Correlations with VQM, along with averages and deviation values.

The first set of results is included in Table~\ref{tab:random} (first row) and depicted in Figure~\ref{fig:randomregval8020}. We notice a definite improvement compared to blind prediction (Table~\ref{tab:NR1video} and Figure~\ref{fig:vqmvsm13videos}). If we exclude SVR, that has the smallest correlation to VQM ($63\%\pm3$), all other prediction algorithms are consistently accurate, in terms of both correlations to VQM (in the $78-97\%$ range) and deviations (in the $0.4-6\%$ range). Even more remarkably, all our prediction-based metrics work on the whole range of network conditions (0-10\% packet loss) and bitrates (64kbps to 5.12Mbps). We can confidently claim so thanks to the low deviations reached when averaging across all network conditions (0.4-6\% range).

\section{Performance vs size of the training dataset}
\label{sec:sizetraining}

Having established the accuracy of prediction-based metrics across a variety of machine learning methods, our next aim was to explore how the size of the training dataset affected the metrics accuracy. In other words, how many video conditions would a service provider have to use to train accurate predictions models?

To this end, we followed the same evaluation method of Section~\ref{sec:evalcom}, splitting the 960-sample dataset in five subsets and performing a 5-fold cross-validation test. However, this time we evaluated the machine learning algorithms on different training sample sizes. Figure~\ref{fig:sizes} and Table~\ref{tab:random} capture all the results, considering training and testing samples of (80\%;20\%), (60\%;40\%), (40\%;60\%) and (20\%;80\%), respectively. As expectable, the reduction of the training set leads to an increase in error. However, this is comparably small. Overall, when the training set is reduced from 80\% to 60\%, 40\% and 20\%, the accuracy drops by an average of 2.4\%, 4.7\% and 7.9\%, respectively. For instance, if we look at our 960-sample dataset we can expect an overall accuracy in the area of 86.6\% (using 768 samples for training), 84.1\% (using 576 samples) and 78.7\% (using 192 samples).

Assessing several machine learning approaches is very useful in pinpointing the most effective algorithms and, in turn, pursue even better performance. For instance, neural networks show a consistent performance in excess of 85\%, even when the training set is reduced down to 20\%. The best performers are the Ensemble Regression Trees, particularly LSB with its 97\% accuracy (with 80\% training samples) that drops only to 93\% (with 20\% training samples). ERT-LSB is also the best performer on blind predictions (91\% overall accuracy, Table~\ref{tab:NR1video}), which makes this the algorithm of preference for our predictive NR method.

\begin{table*}[t!]
\centering
\caption{Overall performance of nine machine learning algorithms, for different sizes of the training and testing datasets. Values indicate overall PCC correlations to VQM (and standard deviations). Cell colors give qualitative correlation levels: green (best); orange (median); and red (worst).}
 \begin{tabular}{|c|c|c|c|c|c|c|c|c|c|}
 \hline
 TR/TE&LR&RT&ERT-LSB&ERT-BR&EDT-AB&GPR&SVR&FNN&CNN\\
 \hline
 80/20&0.782041&0.933185&{\cellcolor{OliveGreen}0.970729}&0.947424&0.829383&{\cellcolor{Yellow}0.888107}&{\cellcolor{Maroon}0.633710}&0.922541&0.886735\\
 &$\pm$0.065142&$\pm$0.011086&{\cellcolor{OliveGreen}$\pm$0.004461}&$\pm$0.006286&$\pm$0.038664&{\cellcolor{Yellow}$\pm$0.024727}&{\cellcolor{Maroon}$\pm$0.038550}&$\pm$0.043051&$\pm$0.151904\\
 \hline
 60/40&0.773775&0.925334&{\cellcolor{OliveGreen}0.965980}&0.936085&0.783163&0.824391&{\cellcolor{Maroon}0.615687}&0.928309&{\cellcolor{Yellow}0.824829}\\
 &$\pm$0.038017&$\pm$0.004382&{\cellcolor{OliveGreen}$\pm$0.002612}&$\pm$0.006419&$\pm$0.068068&$\pm$0.041317&{\cellcolor{Maroon}$\pm$0.011983}&$\pm$0.047652&{\cellcolor{Yellow}$\pm$0.191833}\\
 \hline
 40/60&0.769667&0.901927&{\cellcolor{OliveGreen}0.956091}&0.917091&{\cellcolor{Yellow}0.799145}&0.778152&{\cellcolor{Maroon}0.573457}&0.922428&0.762948\\
&$\pm$0.033387&$\pm$0.009821&{\cellcolor{OliveGreen}$\pm$0.002103}&$\pm$0.010665&{\cellcolor{Yellow}$\pm$0.037741}&$\pm$0.018963&{\cellcolor{Maroon}$\pm$0.011812}&$\pm$0.015941&$\pm$0.119596\\
 \hline
20/80&0.765240&{\cellcolor{Yellow}0.849167}&{\cellcolor{OliveGreen}0.928413}&0.875535&0.740936&0.750394&{\cellcolor{Maroon}0.469481}&0.862465&0.852633\\
&$\pm$0.025263&{\cellcolor{Yellow}$\pm$0.023107}&{\cellcolor{OliveGreen}$\pm$0.008783}&$\pm$0.012865&$\pm$0.068015&$\pm$0.018335&{\cellcolor{Maroon}$\pm$0.026172}&$\pm$0.040414&$\pm$0.023079\\
 \hline
 \end{tabular}
  \label{tab:random}
\end{table*}

\begin{table*}[t!]
\centering
\caption{Overall computational time (in seconds) for the training of nine machine learning algorithms, for different sizes of the training and testing datasets.  Cell colors give completion performance: green (best); orange (median); and red (worst).
}
 \begin{tabular}{|c|c|c|c|c|c|c|c|c|c|}
 \hline
 TR/TE&LR&RT&ERT-LSB&ERT-BR&EDT-AB&GPR&SVR&FNN&CNN\\
 \hline
 80/20&{\cellcolor{OliveGreen}0.000247}&0.011658&4.482096&3.633728&{\cellcolor{Maroon}98.460782}&3.637884&0.030279&{\cellcolor{Yellow}0.445022}&0.441135\\
 &{\cellcolor{OliveGreen}$\pm$0.000057}&$\pm$0.00011&$\pm$0.023245&$\pm$0.007009&{\cellcolor{Maroon}$\pm$16.802947}&$\pm$2.320699&0.001663&{\cellcolor{Yellow}$\pm$0.033905}&$\pm$0.027306\\
 \hline
 60/40&{\cellcolor{OliveGreen}0.000187}&0.010016&4.045408&4.096745&{\cellcolor{Maroon}51.241541}&1.806842&0.016445&0.379740&{\cellcolor{Yellow}0.384112}\\
 &{\cellcolor{OliveGreen}$\pm$0.000028}&$\pm$0.000652&$\pm$0.007866&$\pm$0.091037&{\cellcolor{Maroon}$\pm$9.485821}&$\pm$1.052901&$\pm$0.000839&$\pm$0.009432&{\cellcolor{Yellow}$\pm$0.031723}\\
 \hline
 40/60&{\cellcolor{OliveGreen}0.000123}&0.008180&3.639312&3.232984&{\cellcolor{Maroon}20.608864}&0.269672&0.007259&0.367064&{\cellcolor{Yellow}0.359321}\\
&{\cellcolor{OliveGreen}$\pm$0.000015}&$\pm$0.000134&$\pm$0.010171&$\pm$0.001575&{\cellcolor{Maroon}$\pm$2.474752}&$\pm$0.035411&$\pm$0.000528&$\pm$0.012446&{\cellcolor{Yellow}$\pm$0.00516}\\
 \hline
20/80&{\cellcolor{OliveGreen}0.000078}&0.006863&3.245447&3.040967&{\cellcolor{Maroon}5.402421}&0.066810&0.001883&{\cellcolor{Yellow}0.32525}&0.319859\\
&{\cellcolor{OliveGreen}$\pm$0.000007}&$\pm$0.000059&$\pm$0.004206&$\pm$0.005620&{\cellcolor{Maroon}$\pm$0.156247}&$\pm$0.009270&$\pm$0.000149&{\cellcolor{Yellow}$\pm$0.007258}&$\pm$0.011671\\
 \hline
 \end{tabular}
  \label{tab:times}
\end{table*}

\section{Computational Trade-offs}
\label{sec:comptimes}

The models of our performance metric are trained in the background (Figure~\ref{fig:serverclient}, top), before being used in the client (Figure~\ref{fig:serverclient}, bottom). Thus, the running time of the learning algorithms will not affect the real-time quality metric computational times. Still, it is interesting to see the trade-offs achievable with the different machine learning techniques, as these will affect the service provider’s ability to manage video datasets at scale. 

To this end, we follow the same evaluation method of Section~\ref{sec:sizetraining}, splitting the 960-sample dataset in five subsets, performing a 5-fold cross-validation test and evaluating the algorithms on the four training-testing subdivisions considered earlier,  i.e. 80\%-20\%; 60\%-40\%; 40\%-60\%; 20\%-80\%. In each case, we measure the time incurred to train the model. We perform this process on a Laptop (HP EliteBook) with an Intel Core i7 processor and 7,7GB of RAM memory. 

As it could be expected, smaller training sets incur faster completion times (Table~\ref{tab:times}). However, the difference is not significant (computation time orders do not vary between the 20/80 split and the 80/20 split). The fastest algorithm was LR, with computation time in the millisecond scale. Interestingly, this is not the least accurate metric (overall 77\% correlation, Table~\ref{tab:random}). 

On the other end of the range, the ensemble regression trees (ERT-BR and ERT-LSB) incur times ranging from 3 to 4.5 seconds. This is because they have to build 500 consecutive models before they can complete the trained models. Yet, these lead to the most significant accuracy.

Even when deployed in on a low-spec laptop, the computational times of the prediction metrics are negligible and, certainly, compatible with the typical background processes of a service provider. Also, in a commercial setting the background QoE processes will be supported by dedicated servers and, when necessary, data centers or cloud services. Hence, the times involved in characterizing the video dataset would not constitute a bottleneck.

Figure~\ref{fig:timesbest} shows the performance trade-offs (accuracy and computational time) of the different learning algorithms (colored symbols), considering all training/testing combinations. To benchmark our metrics, we include also SSIM and VQM (in black).  With the exception of SVR, all other learning models perform much better than SSIM, although the former are NR and the latter is FR. Of crucial importance is our finding of the learning computational times, which are four orders smaller than SSIM and six orders smaller than VQM. Thus prediction metrics are comparably as accurate as VQM while scaling significantly better.

\section{Related work on Machine Learning for NR Quality Assessment}
\label{sec:background}

In our previous research, we have conducted a range of preliminary studies that have provided basis and motivations to the present paper. Our most relevant works are summarized next. Our earlier attempts to develop NR metrics based on conventional features (i.e. without using machine learning), lead to a formula that combined scene complexity and motion and could be computed in real-time~\cite{Liotta:2013:IVQ:2536853.2536903}. At the same time, we were exploring the use of machine learning to address fundamental limitations of conventional NR metrics~\cite{TorresES2016}, mainly the lack of generality and poor performance. In~\cite{vladohttp} we showed the use of Reinforcement Learning to optimize video quality in adaptive streaming, without using complex heuristics. In~\cite{TorresNRQoE2015} we showed how artificial neural networks could determine a linear combination of blur and noise that performed significantly better than these two NR metrics in isolation. Finally, our recent survey of machine learning in NR video quality assessment~\cite{decoqoedtv} provides a snapshot of the state-of-the-art on which our work is based. A selection of the most relevant on-going efforts is briefly described below.  

In the last decade, several researchers have explored the machine learning path in order to improve both the generality and accuracy of NR metrics. Already in 2002, Gastaldo et al. introduced one of the first methods to estimate the video quality using artificial neural networks~\cite{gastaldoobjective2002}. They proposed the use of circular back propagation networks (based on bitstream layer parameters) in order to mimic the users perception of compressed MPEG2 videos. Their approach showed promising results on a 12-video dataset from the motion picture expert group (MPEG). Their study focused on video distortions deriving merely from compression and explored a specific machine learning method.

Also working on compressed videos, Le Callet et al.~\cite{lecalletconvolutional2006} employed an interesting convolutional neural network as a Reduced Reference (RR) method to allow a continuous-time quality estimation and scoring of the video. Unlike our NR approach, in which the server transmits the machine learning model updates only on service launch and in the case that an update is due, their method (as any RR metric) requires the transmission of features extracted from the original video together with the video under scrutiny.

Zhu et al.~\cite{KongfengZhu2014_CSVT} proposed the use of neural networks and features extracted from the analysis of Discrete Cosine Transform (DCT) coefficients of each decoded frame from a video sequence to predict its quality. Their approach showed good correlation results in compressed videos of four different well-known datasets. However, their method is distortion specific, and thus of a more limited scope than our case. Furthermore, the complexity of the approach makes it not viable to real-time deployments.

Staelens et al.~{\cite{Staelens14}  presented an NR video quality estimation method which uses a symbolic regression framework trained on a large set of parameters extracted from the codec. While obtaining good correlation with subjective tests, their approach is suited only to H.264 compressed streams, thus loosing on generality. 

Similar principles were proposed in~\cite{Sogaard15} by using features extracted from specific codecs (MPEG or H.264/AVC), the analysis of DCT coefficients, the estimation of the quantization level used in the I-frames to measure quality of videos distorted by only the compression process. They show high correlation with some state-of-the-art metrics (FR, RR and NR). However, their approach is only suited to a specific type of codec and the complexity of the feature extraction process makes this NR metric incompatible with real-time applications. 

Shahid et al.~\cite{DBLP:conf/qomex/ShahidPWBL15} proposed a model combining different bitstream-layer features using an Artificial Neural Network to estimate the quality. They tested their method on compressed videos but focused on correlations with PSNR.


The key differentiator between our work and other valuable on-going efforts is our focus on a generic learning framework for assessing end-to-end streaming in real-time. Our predictive method (Figure~\ref{fig:serverclient}) and evaluation methodology (Figure~\ref{fig:evmethod}) are completely independent from type of video, compression, benchmarking quality, transmission means and machine learning algorithm. We place the heavy part of the machine learning (training) on a background process, allowing for a light-weight evaluation metric to be executed in real-time, even on thin clients. We do not have to rely on synthetic impairments and have a system that can be employed in a typical video service provisioning platform.

\section{Conclusion}
\label{sec:conclusion}

No-Reference (NR) video quality methods have the potential to provide real-time video quality assessment and automated quality control, in situations in which traditional subjective studies or FR methods are unfeasible, due to their high complexity and time requirements. However, as we demonstrated in our previous work~\cite{TorresES2016}, classic NR fails to deliver accurate results over broad operational conditions and, specifically, cannot handle network-impaired streams. On the other hand, the more advanced NR methods (e.g. based on machine learning tend to be heavyweight and often lack generality).

In this work, we introduce a generic machine learning framework (Figure~\ref{fig:serverclient}) that allows deriving a predictive NR assessment metric. We explored the efficiency and accuracy of our metric for a broad representation of supervised-learning techniques (Table~\ref{tab:params}), using a varied video dataset (Table~\ref{tab:database}). 

Through an extensive analysis (Section~\ref{sec:evalworst} to~\ref{sec:comptimes}), we demonstrated that our approach is not tied to any particular type of video, compression, or transmission means. In fact, the metric performance remains remarkably high when the training set is reduced from 80 to 20\% (Table~\ref{tab:random}), indicating that models can accurately predict 80\% of unknown conditions.

We are particularly keen to have developed an NR metric that operates accurately under lossy networks. We tested the whole 0-10\% packet loss range, which reflects the most extreme Internet conditions. Overall, we have achieved an over 97\% correlation to VQM, demonstrating that it is possible to develop an NR metric that is as accurate as an FR method, while allowing real-time assessment of video quality in realistic streaming scenarios.


%

\appendices

\section*{Acknowledgment}
This work has been carried out in the context of the European Research Council project BROWSE (Beam-steered Reconfigurable Optical-Wireless System for Energy-efficient communication - Grant 291632) and the ICT COST Action 3D-ConTourNet (IC1105).

\ifCLASSOPTIONcaptionsoff
  \newpage
\fi



\bibliographystyle{myIEEEtran}

\bibliography{double}
\balance

%

%

\end{document}